\documentclass[twocolumn, times]{aastex63}

\usepackage{amsmath,verbatim, nicefrac,color, soul}
\usepackage{natbib}

\usepackage[caption=false]{subfig}

\newcommand\Hl[1]{\colorbox{yellow}}

\shorttitle{Multi-Epoch Scintillation Studies}
\shortauthors{J. E. Turner \lowercase{et al}.}

\begin{document}
\title{The Pulsar Science Collaboratory: Multi-Epoch Scintillation Studies of Pulsars}

\author[0000-0002-2451-7288]{Jacob E. Turner}
\affiliation{Green Bank Observatory, P.O. Box 2, Green Bank, WV 24944, USA}

\author[0000-0001-8233-5442]{Juan G. Lebron Medina}
\affiliation{Department of Physics, 
University of Puerto Rico at Mayagüez, 6V66+C8C, Mayagüez, 00680, Puerto Rico}

\author[0000-0001-6620-5752]{Zachary Zelensky}
\affiliation{Department of Physics, Pennsylvania State University, State College, PA 16801}

\author[0000-0001-8344-7235]{Kathleen A. Gustavson}
\affiliation{Nicolet High School, 6701 North Jean Nicolet Road, Glendale, WI 53217}

\author[0000-0002-9578-9221]{Jeffrey Marx}
\affiliation{Department of Physics, McDaniel College, 2 College Hill Westminster, MD, 21157}

\author[0009-0007-3378-0508]{Manvith Kothapalli}
\affiliation{Northshore Networks, 18101 Avondale Rd NE, Woodinville, WA 98077}

\author[0000-0001-9356-9738]{Luis D. Cruz Vega}
\affiliation{Department of Physics, 
University of Puerto Rico at Mayagüez, 6V66+C8C, Mayagüez, 00680, Puerto Rico}

\author[0000-0003-4560-9920]{Alexander Lee}
\affiliation{Department of Aeronautics \& Astronautics, University of Washington, Guggenheim Hall 211,3940 Benton Lane NE,UWAA Box 352400, Seattle, WA 98195-2400}

\author[0000-0002-9951-5519]{Caryelis B. Figueroa}
\affiliation{Department of Physics, 
University of Puerto Rico at Mayagüez, 6V66+C8C, Mayagüez, 00680, Puerto Rico}

\author{Daniel E. Reichart}
\affiliation{Skynet Robotic Telescope Network,University of North Carolina, Chapel Hill, NC 27599, USA}

\author{Joshua B. Haislip}
\affiliation{Skynet Robotic Telescope Network,University of North Carolina, Chapel Hill, NC 27599, USA}

\author{Vladimir V. Kouprianov}
\affiliation{Skynet Robotic Telescope Network,University of North Carolina, Chapel Hill, NC 27599, USA}

\author{Steve White}
\affiliation{Green Bank Observatory, P.O. Box 2, Green Bank, WV 24944, USA}

\author{Frank Ghigo}
\affiliation{Green Bank Observatory, P.O. Box 2, Green Bank, WV 24944, USA}

\author{Sue Ann Heatherly}
\affiliation{Green Bank Observatory, P.O. Box 2, Green Bank, WV 24944, USA}

\author[0000-0001-7697-7422]{Maura A. McLaughlin}
\affiliation{Department of Physics and Astronomy, West Virginia University, P.O. Box 6315, Morgantown, WV 26506, USA}
\affiliation{Center for Gravitational Waves and Cosmology, West Virginia University, Chestnut Ridge Research Building, Morgantown, WV 26505, USA}

\begin{abstract}
We report on findings from scintillation analyses using high-cadence observations of eight canonical pulsars with observing baselines ranging from one to three years. We obtain scintillation bandwidth and timescale measurements for all pulsars in our survey, scintillation arc curvature measurements for four, and detect multiple arcs for two. We find evidence of a previously undocumented scattering screen along the line of sight (LOS) to PSR J1645$-$0317, as well as evidence that a scattering screen along the LOS to PSR J2313$+$4253 may reside somewhere within the Milky Way's Orion-Cygnus arm. We report evidence of a significant change in the scintillation pattern in PSR J2022$+$5154 from the previous two decades of literature, wherein both the scintillation bandwidth and timescale decreased by an order of magnitude relative to earlier observations at the same frequencies, potentially as a result of a different screen dominating the observed scattering. By augmenting the results of previous studies, we find general agreement with estimations of scattering delays from pulsar observations and predictions by the NE2001 electron density model but not for the newest data we have collected, providing some evidence of changes in the ISM along various LOSs over the timespans considered. In a similar manner, we find additional evidence of a correlation between a pulsar's dispersion measure and the overall variability of its scattering delays over time. The plethora of interesting science obtained through these observations demonstrates the capabilities of the Green Bank Observatory's 20m telescope to contribute to pulsar-based studies of the interstellar medium.
\end{abstract}

\section{Introduction}\label{intro}
Pulsars provide a valuable lens through which to study the structure and behavior of the ionized interstellar medium (ISM). When pulsar radio emission traveling towards Earth interacts with free electrons in the ISM, the resulting phase shifts in the emission, along with the multipath propagation of the signal, results in time and frequency-evolving interference patterns, seen in pulsars' dynamic spectra, when the signal recombines at our telescopes. The regions of constructive interference in these spectra, known as scintles, can provide insight into the electron density fluctuations of the ISM along a given line of sight (LOS). Additionally, tracking these scintillation patterns over long periods of time can also allows us to monitor the evolution of these LOSs on months-to-years-long timescales.
\par If scintles are sufficiently resolved in both observing frequency and time, they can give rise to parabolic features, known as scintillation arcs \citep{OG_arcs}, in the 2D Fourier transform of a dynamic spectrum, also known as the secondary spectrum. These arcs are thought to originate from localized screens (AU-scale) that dominate the scattering the pulsar emission experiences along its propagation path. Additionally, by using the curvature of these arcs along with the corresponding pulsar's distance and LOS velocity across that screen, $\textbf{V}_{\textrm{eff}, \perp}$, we can constrain the screen's location along the LOS \citep{OG_arcs, walker_2004, cordes_2006_refraction, stine_survey}. 
\par Studying this scintillation phenomenon is the primary goal of the pulsar scintillation group within the Pulsar Science Collaboratory (PSC, formerly Pulsar Search Collaboratory \citep{psc_paper}), which is an organization founded in 2007 to provide research opportunities to high school and undergraduate students by means of looking through Green Bank Telescope (GBT) drift scan data for evidence of new pulsars. The program has since expanded to include a number of other research projects that use the Green Bank Observatory's 20m telescope, including studies of giant pulses, pulsar timing, and pulsar scintillation. 
\par While not as sensitive as the GBT, the 20m's primary function as an outreach instrument means that observing proposals are not required for usage, allowing for high-cadence, years-long campaigns, wherein each observation may last upwards of two hours, that may not have received time had similar observing campaigns been requested on a larger telescope. The flexibility offered by the 20m can allow for observing setups with such a breadth and density of data that they can rival, and in the case of the lengths of individual observations, surpass, the times of timing campaigns typically used for pulsar timing array science.
\par High-cadence, long observations are ideal for pulsar studies of dynamic systems such as the ISM, where vastly different phenomena can be observed on hour, week, and year-long timescales. Observations on the order of multiple hours are necessary to measure scintillation timescales, which provide important constraints on pulsar velocities when these velocities are not measurable through proper motion. Alternatively, if velocities are measurable through proper motion, measuring scintillation timescales provides important constraints on scattering screens along the line of sight. These long observations are also important for more accurate measurements of scintillation bandwidths, as one can average over more scintles. They are also crucial for developing more accurate ISM models that can better predict observable quantities such as scattering delays, critical for developing high-precision pulsar timing arrays. Additionally, fully visualizing scintles in both frequency and time in a given observation is a necessity for tracking phenomena like scintillation arcs. Observing cadences on the order of a week to a month are valuable for tracking ISM evolution over a pulsar's refractive timescale, and necessary for tracking the structure of transient features in data such as extreme scattering events. Making such observations with baselines of many years results in highly-detailed monitoring of the ISM, allowing one to significantly constrain distances to scattering screens, as well as track changes in scattering screen distances and/or properties with time.
\par In this paper we report findings from the PSC pulsar scintillation group based on 2.5 years of observations of eight canonical pulsars. In Section \ref{data}, we discuss details regarding our data. In Section \ref{analyses}, we describe our data processing procedure and our various scintillation analyses. In Section \ref{results}, we discuss the results of our analyses, first detailing general population trends and then describing intriguing features found along the LOSs of specific pulsars. Finally, in Section \ref{concl}, we discuss both our conclusions and plans for follow-up observations using both the telescope we currently use for observations as well as more sensitive instruments. 
\section{Data}\label{data}
Our observations were taken using the 20m telescope at the Green Bank Observatory using the ``all" filter, which observes from 1350$-$1750 MHz. This filter was chosen because it  offers the widest possible observing bandwidth available on the 20m, allowing for tighter constraints on scintillation parameter estimates due the greater number of scintles across the observing band, as well as allowing for the study of scintillation evolution over observing frequency. Such studies can inform our understanding of the turbulence of the ISM and whether this turbulence may adhere to a Kolmogorov wavenumber spectrum. To accommodate the limited sensitivity of the 20m, we observed only pulsars with flux densities greater than 10 mJy at 1400 MHz. Additionally, in order to ensure detection of multiple scintles, we further limited our study to pulsars with predicted or measured average scintillation bandwidth upper limits less than 100 MHz. These values were determined using predictions from the NE2001 electron density model \citep{NE2001} at 1400 MHz, as well as existing literature values, when available. To achieve sufficient coverage of the two longest scientific timescales described in Section \ref{intro}, each pulsar was observed with at least a monthly cadence, and occasionally up to a weekly cadence, with our total data set for all pulsars spanning MJDs 59382$-$60384.  Additionally, while early on in our campaign, observations of each pulsar lasted 30 minutes, to properly measure scintillation timescales for the reasons described in Section \ref{intro}, this was increased to 120 minutes in the final year of our observations. Observations were taken with a sampling time of 4.18 ms, sufficient resolution for the long-period pulsars we observed, and ephemerides were acquired from the Australia National Telescope Facility (ATNF) pulsar catalog. After data was processed, RFI excision was performed both by means of a custom median-smoothed zapping script, as well as manual removal of individual pixels.

\par After accounting for radio frequency interference (RFI) and various filters that are automatically applied by the observation software, our data typically spanned around 170 MHz in observing bandwidth centered around 1435 MHz, although in practice our usable observing bandwidth and center frequency varied by a few tens of MHz depending on the RFI environment in a given observation. Depending on the S/N (signal-to-noise) of each observation, after folding and summing over polarizations, our data was averaged in frequency to yield either 1 or 0.5 MHz channels (256 and 512 frequency channels, respectively, across the original 400 MHz observing band) and averaged in time to yield either 10 or 20 second subintegrations. The lower resolution options were chosen if scintillation structures were visible, but not bright enough for our subsequent analyses. To ensure we could guarantee at least three frequency channels per scintle \citep{turner_scat}, these frequency resolution constraints further limited the sources in our study to pulsars with predicted or measured scintillation bandwidth lower limits greater 1.5 MHz for the higher frequency resolution data and 3 MHz for the lower resolution data. 
\section{Analyses}\label{analyses}
To examine the effects of scintillation in our data, for each observation, we used \textsc{pypulse} \citep{pypulse} to extract its dynamic spectrum, defined as
\begin{equation}
\label{dynspec}
S(\nu,t)=\frac{P_{\rm{on}}(\nu,t)-P_{\rm{off}}(\nu,t)}{P_{\rm{bandpass}}(\nu,t)},
\end{equation}
where $S$ is the observed pulsar signal's intensity at each given observing frequency, $\nu$, and time, $t$, ${P}_{{\rm{bandpass}}}$ indicating the observation's total power, and ${P}_{{\rm{on}}}$ and ${P}_{{\rm{off}}}$ indicating the power in all on- and off-pulse components, respectively, at a given frequency and time. The on-pulse window was defined using the reference pulse created by the template-generating algorithm in \textsc{pypulse}'s SinglePulse set of functions. The secondary spectrum for each observation was then created by taking the squared modulus of the two-dimensional Fourier transform of that dynamic spectrum. 
\subsection{Scintillation Bandwidths \& Timescales}
To determine a given dynamic spectrum's characteristic scintle widths in frequency and time, we extracted its 2D autocorrelation function (ACF) using \textsc{pypulse} \citep{pypulse}. Subsequently, we took summations of this 2D ACF over time and frequency to create 1D time and frequency ACFs, respectively. The scintillation bandwidth for a given subband was then estimated by fitting either a Lorentzian or a Gaussian to the frequency ACF and finding the half width at half maximum (HWHM) of the frequency ACF fit. Similarly, the scintillation timescale was estimated by fitting either a Gaussian or a Lorentzian to the time ACF and finding the half width at $1/e$ of the time ACF fit \citep{Cordes_1985}. We attempted to measure the $1/e$ of our ACF fits by finding roots using scipy's brentq algorithm. In the event that the algorithm did not converge, we used the HWHM instead.
\par
\begin{figure}[t]
\centering
\hspace*{-0.5cm}
\includegraphics[scale = 0.65]{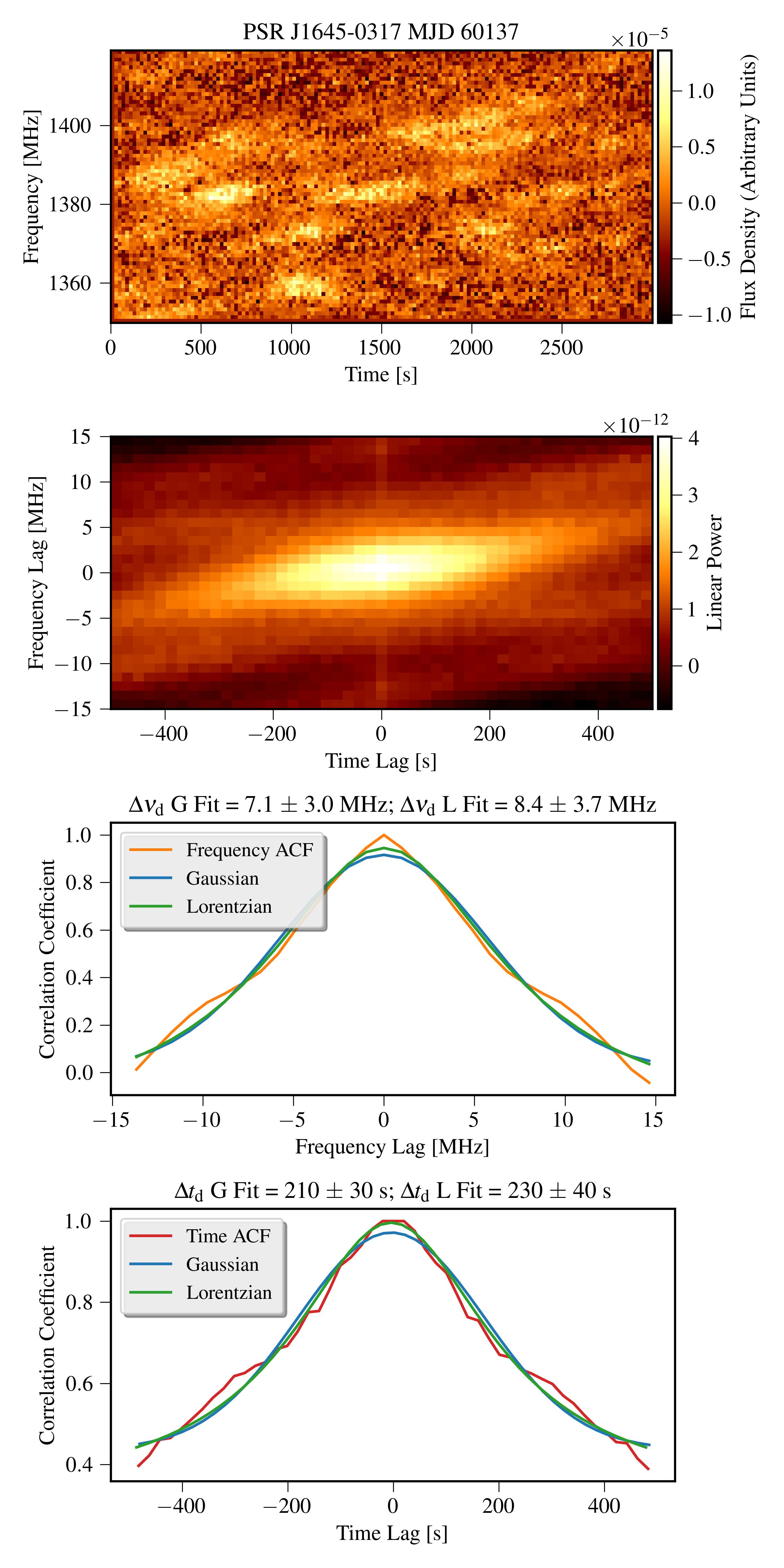}
\caption{(a) A dynamic spectrum for PSR J1645--0317 on MJD 60137. (b) The corresponding 2D autocorrelation function. (c) The 1D frequency slice (orange) of the 2D autocorrelation function with corresponding fits. (d) The 1D time slice of the 2D autocorrelation function (red) with its corresponding fits.}
\label{ex_acf_fits}
\end{figure}

 As mentioned in Section \ref{data}, as a consequence of the variability that RFI introduces into our data, the center frequencies in our usable dynamic spectra can vary by a few tens of MHz from epoch to epoch. To compensate, for all analysis and visualizations that use multiple measurements of the above quantities, such as weighed averages and timeseries, we first scale all measurements to their equivalent values at 1400 MHz assuming their frequency evolution behaves as expected for a medium with a Kolmogorov wavenumber spectrum, i.e., $\propto\nu^{4.4}$ for scintillation bandwidth and $\propto\nu^{1.2}$ for scintillation timescale. While this may introduce some additional source of error into our results, as it is not guaranteed that the LOS to a given pulsar will yield these power laws for our observables \citep{Levin_Scat,turner_scat,epta_scint,Turner_2024}, we believe this should result in minimal effects on our results due to the relatively small frequency range. An example dynamic spectrum and its corresponding 2D ACF and frequency and time ACFs along with their fits are shown in Figure \ref{ex_acf_fits}.


\par It is important to note that, in the majority of studies that use an ACF analysis to determine scintillation parameters, scintillation bandwidths are fit using Gaussians. However, given that the ISM's pulse broadening function, a time domain function, is characterized by a one-sided decaying exponential, we fit the corresponding frequency domain ACF using a Lorentzian, as the two functions are Fourier pairs. To maintain consistency with the majority of existing scintillation studies though, we performed both Gaussian and Lorentzian fits on our ACFs.
\par The precision in our estimations of these scintillation parameters was limited by the number of scintles visible in a given observation. Consequently, we found our uncertainties dominated by this finite scintle effect, which is defined as 
\begin{equation}
\label{finite_scintle}
\begin{split}
\epsilon &  = \frac{\Delta \nu_{\rm d}}{2\ln(2) N_{\rm scint}^{1/2}} \\
& \approx \frac{\Delta \nu_{\rm{d}}}{2\ln(2)[(1+\eta_{\text{t}}T/\Delta t_\text{d})(1+\eta_\nu B/\Delta \nu_\text{d})]^{1/2}},
\end{split}
\end{equation}
where ${N}_{{\rm{scint}}}$ is the number of scintles seen in the corresponding observation, $T$ and $B$ are the total integration time and total bandwidth, respectively, and ${\eta }_{{\rm{t}}}$ and ${\eta}_{\nu}$ are filling factors that can range from $0.1$ to $0.3$ depending on how one defines the characteristic bandwidth and  timescale, which for this paper are both set to 0.2 \citep{Cordes1986}. The scintillation timescale's finite scintle error is identical except for the replacement of $\Delta \nu_{\rm d}$ with $\Delta t_{\rm d}$ in the numerator and the absence of the factor of $\ln(2)$ in the denominator.

\par Once we estimated scintillation bandwidths and timescales, we then used these quantities to estimate each pulsar's refractive timescale, $t_r$, which, per \cite{stinecon}, can be approximated as

\begin{equation}
    t_r\approx\frac{4}{\pi}\Big(\frac{\nu\Delta t_{\rm d}}{\Delta \nu_{\rm d}}\Big).
\end{equation}

\par Finally, under the assumption that a pulsar's transverse velocity as determined from proper motion, $V_{\rm pm}$, should be equal to the velocity of the ISM along the LOS to the pulsar \citep{arm_thickness,space_velo_1}, $V_{\rm ISS}$, we can estimate the distance to the scattering screen that dominates the scintillation pattern seen in its dynamic spectrum. Merging the expressions from \cite{gupta_velo} and \cite{Cordes_1998}, $V_{\rm ISS}$ is given by
\begin{equation}
\label{iss_velo}
V_{\textrm{ISS}}=A_{\textrm{ISS}}\frac{\sqrt{\Delta \nu_{\textrm{d,MHz}}D_{\textrm{p,kpc}}x}}{\nu_{\textrm{GHz}}\Delta t_{\textrm{d,s}}},
\end{equation}
where $D_{\textrm{p,kpc}}$ is the distance to the pulsar in kpc, $A_{\textrm{ISS}}$ is a factor that depends on various assumptions related to the geometry and uniformity of the medium, and $x=D_{s}/(D_p-D_s)$, where $D_s$ is the distance from the observer to the screen. Note that $x=1$ in the scenario in which the screen is halfway between the pulsar and the observer. In our case, we assume a thin screen and a Kolmogorov medium as in \cite{Cordes_1998}, resulting in $A_{\textrm{ISS}} = 2.53 \times 10^4 \ \textrm{km s}^{-1}$. To determine $D_s$, we set the right hand side of Equation \ref{iss_velo} equal to $V_{\rm pm}$ and solve for $x$, which results in an estimation of $D_s$, since both $D_p$ and $V_{\rm pm}$ are known either through VLBI, pulsar timing, or both. In particular, we get our distances and velocities from \citet{chatterjee_2001}, \cite{brisken_2002}, \cite{Chatterjee_2009}, and \cite{Deller_2019}. Additionally, we assume the screen is isotropic, we ignore orbital velocities for binary pulsars, and we ignore the Earth's velocity for all pulsars.

\subsection{Scintillation Arcs}
\begin{figure}[t]
\centering
\includegraphics[scale = 0.45]{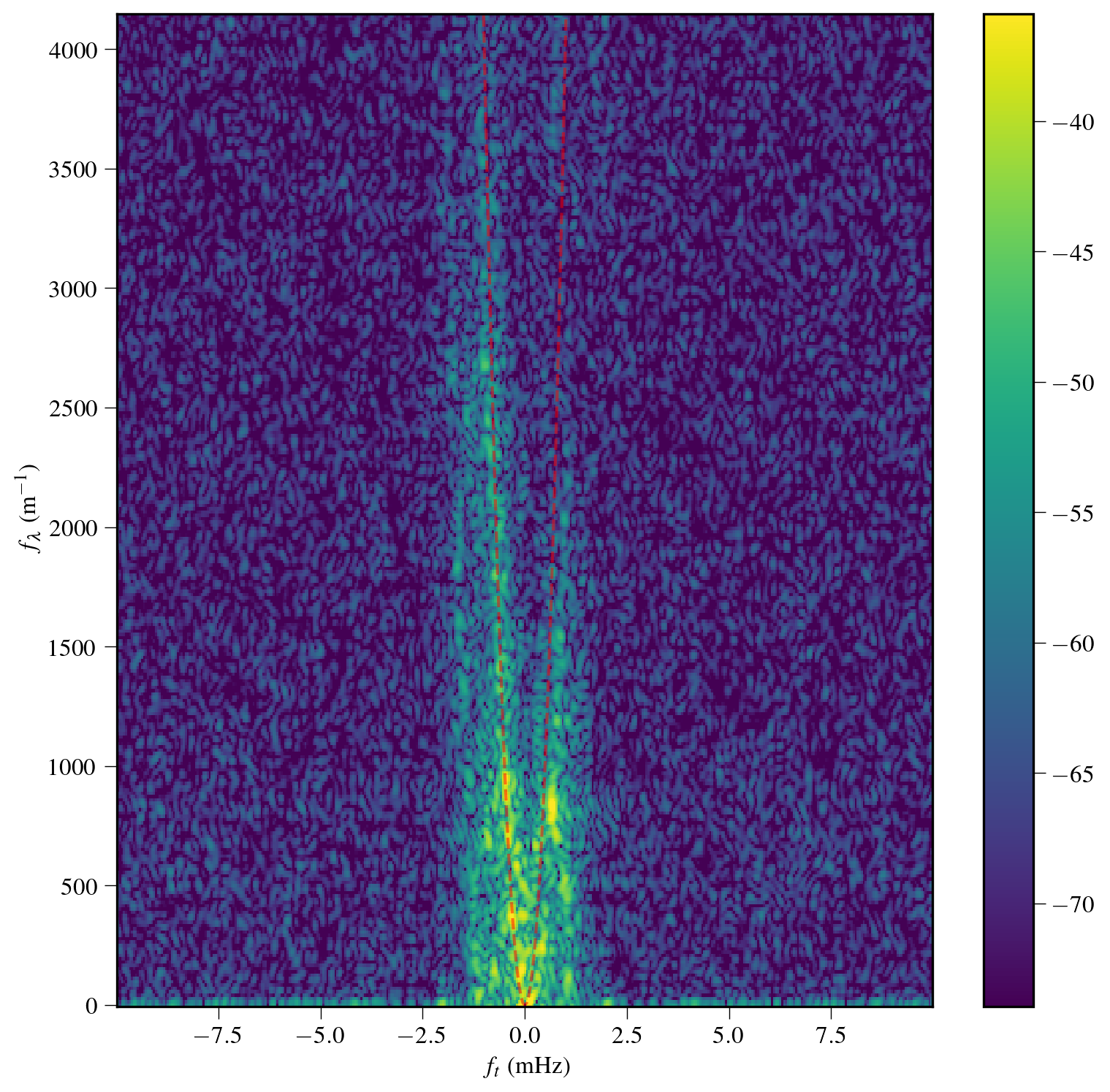}
\caption{An example scintillation arc with its corresponding fit (dashed red line) for PSR J2022+5154 on MJD 60306. The color bar is in units of log power (dB).}
\label{ex_arc_fit}
\end{figure}

After creating secondary spectra, we determined arc curvatures, $\eta$, for all visible arcs using the Hough transform technique implemented in the \textsc{scintools} package \citep{Reardon_2020}. An example arc and its associated fit can be seen in Figure \ref{ex_arc_fit}. 

We then estimated the screen distance associated with each arc via

\begin{equation}
    \eta = \frac{\lambda^2}{2c}\frac{D_{\rm eff}}{(|V_{\rm eff}|\cos\alpha)^2},
\end{equation}
where $\lambda$ is the observing wavelength and $D_{\rm eff}$ is the effective distance to the pulsar, defined as $D_p D_s/(D_p-D_s)$ \citep{McKee_2022}. Additionally, $V_{\rm eff}$ is the effective velocity of the pulsar, defined as 
\begin{equation}
    V_{\textrm{eff}} =\frac{1}{s}V_{\textrm{ISS}} -\frac{1-s}{s}V_{\textrm{pulsar}}-V_{\textrm{Earth}},
\end{equation}
where $s=1-D_s/D_p$, and $\alpha$ is the angular orientation of the screen, defined as between the major axis of the image on the sky and the system's effective velocity \citep{OG_arcs,cordes_2006_refraction,Reardon_2020,McKee_2022}. 
\par For simplicity, we assume the pulsar velocity dominates the system. Furthermore, we do not know screen orientation $\alpha$ for any of the pulsars in our study, and so make the simplistic assumption of a screen orientation parallel to the path of the pulsar, i.e., $\alpha =0$. Consequently,  our screen distance values should be treated as lower limits. Additionally, more detailed screen distance analyses would account for the velocities inherent in all aspects of the system, including the pulsar, ISM, orbit of the Earth, and orbit of the pulsar if it is in a binary, and track the variations in curvature over many observations to more precisely constrain screen distances \citep{Reardon_2020}. However, both our measurement precision and observing baselines are insufficient for this level of analysis, and so we represent our screen distance estimations as weighted averages of the measurements of a given arc from all epochs. 

\section{Results}\label{results}
\begin{deluxetable*}{CCCCCCCCCC}

\tablewidth{0pt}
\tablecolumns{10}

\tablecaption{Weighted Average Scintillation Parameters \label{scint_table}}
\tablehead{ \multicolumn{2}{C}{\text{Pulsar}} & \colhead{$\overline{\Delta \nu_{\text{d;1400,G}}}$} & \colhead{$\overline{\Delta \nu_{\text{d;1400,L}}}$} & \colhead{$N_{\Delta \nu_{\text{d;1400}}}$} & \colhead{$\overline{\Delta t_{\text{d;1400,G}}}$} & \colhead{$\overline{\Delta t_{\text{d;1400,L}}}$} &\colhead{$N_{\Delta t_{\text{r;1400}}}$} & \colhead{$\overline{t_{\text{r;1400,G}}}$} & \colhead{$\overline{t_{\text{r;1400,L}}}$}\\ \colhead{J2000 Epoch} & \colhead{B1950 Epoch} & \colhead{\text{(MHz)}} & \colhead{\text{(MHz)}} & \colhead{} & \colhead{\text{(min)}} & \colhead{\text{(min)}} & \colhead{} & \colhead{\text{(days)}} & \colhead{\text{(days)}}    \vspace{0.05cm}}
\startdata
\text{J}0332{+}5434 & \text{B}0329{+}54 & 2.5 $\pm$0.6 & 2.8 $\pm$0.7 & 32 & 6.1 $\pm$ 0.7 & 7.6 $\pm$ 1.0 & 32 & 3.0 $\pm$ 0.8 & 3.4 $\pm$ 0.9\\
\text{J}0826{+}2637 & \text{B}0823{+}26 & 17.0 $\pm$ 7.8 & 21.8 $\pm$ 11.3 & 6 & 3.1 $\pm$ 0.4 & 3.1 $\pm$ 0.5 & 6 & 0.2 $\pm$ 0.1 & 0.2 $\pm$ 0.1\\
\text{J}0922{+}0638 & \text{B}0919{+}06 & 4.2 $\pm$ 1.4 & 4.4 $\pm$ 1.6 & 4 & 3.0 $\pm$ 0.3 & 3.2 $\pm$ 0.3 & 4 & 0.6 $\pm$ 0.1 & 0.4 $\pm$ 0.1\\
\text{J}1645\text{$--$}0317 & \text{B}1642\text{$--$}03 & 2.4 $\pm$ 0.7 & 2.1 $\pm$ 0.7 & 14 & 1.9 $\pm$ 0.2 &  2.0 $\pm$ 0.2 & 14 & 1.0 $\pm$ 0.3 & 1.2 $\pm$ 0.4\\
\text{J}2018{+}2839 & \text{B}2016{+}28 & 12.2 $\pm$ 5.8 & 16.2 $\pm$ 9.1 & 3 & 37.3 $\pm$ 8.5 & 56.0 $\pm$ 14.9 & 3 & 3.0 $\pm$ 0.9 & 1.5 $\pm$ 0.7\\
\text{J}2022{+}2854 & \text{B}2020{+}28 & 22 $\pm$ 12 & 25 $\pm$ 15 & 16 & 13.1 $\pm$ 3.3 & 14.1 $\pm$ 4.0 & 16 & 0.7 $\pm$ 0.4 & 0.7 $\pm$ 0.5\\
\text{J}2022{+}5154 & \text{B}2021{+}51 & 1.1 $\pm$ 0.2 & 1.1 $\pm$ 0.2 & 6 & 3.2 $\pm$ 0.2 & 3.8 $\pm$ 0.2 & 6 & 3.6 $\pm$ 0.7 & 4.3 $\pm$ 0.8\\
\text{J}2313{+}4253 & \text{B}2310{+}42 & 4.8 $\pm$ 1.6 & 5.2 $\pm$ 1.6 & 14 & 13.0 $\pm$ 2.0 & 16.2 $\pm$ 2.6 & 14 & 3.4 $\pm$ 1.2 & 3.9 $\pm$ 1.3\\
\enddata
\tablecomments{Weighted average and its weighted 1$\sigma$ uncertainty on the scintillation bandwidths, scintillation timescales, transverse velocities, and refractive timescales scaled to 1400 MHz using Gaussian (G) and Lorentzian (L) fits. Columns with $N$ represent the number of measurements made for that quantity.}
\end{deluxetable*}

\begin{deluxetable*}{CCCCCCCC}

\tablewidth{0pt}
\tablecolumns{7}
\tablecaption{Calculated Scattering Screen Distances \label{screen_table}}
\tablehead{ \multicolumn{2}{C}{\text{Pulsar}}& \colhead{$\overline{D_{s,V,G}}$} & \colhead{$\overline{D_{s,V,L}}$} &  \colhead{$\overline{D_{s,\eta,1}}$} & \colhead{$N_{D_{s,\eta,1}}$} & \colhead{$\overline{D_{s,\eta,2}}$} &  \colhead{$N_{D_{s,\eta,2}}$}\\ \colhead{J2000 Epoch} & \colhead{B1950 Epoch} & \colhead{\text{(kpc)}} &  \colhead{\text{(kpc)}} &\colhead{\text{(kpc)}} & \colhead{} & \colhead{\text{(kpc)}} & \colhead{}  \vspace{0.05cm}}
\startdata
\text{J}0332{+}5434 & \text{B}0329{+}54 & 1.3 $\pm$0.1 & 1.4 $\pm$ 0.1 & 0.3 $\pm$ 0.1 & 7 & 0.7 $\pm$ 0.1 & 1\\
\text{J}0826{+}2637 & \text{B}0823{+}26 & 0.3 $\pm$ 0.1 & 0.3 $\pm$ 0.1 & 0.3 $\pm$ 0.1 & 4 & \textrm{---} & \textrm{---}\\
\text{J}0922{+}0638 & \text{B}0919{+}06 & 0.9 $\pm$ 0.1 & 0.9 $\pm$ 0.1 & \textrm{---} & \textrm{---}  & \textrm{---}& \textrm{---}\\
\text{J}1645\text{$--$}0317 & \text{B}1642\text{$--$}03 & 1.1 $\pm$ 0.2 & 1.3 $\pm$ 0.2 & 0.6 $\pm$ 0.1 & 5 & 2.4 $\pm$0.3 & 8\\
\text{J}2018{+}2839 & \text{B}2016{+}28 & 0.5 $\pm$ 0.1 & 0.6 $\pm$ 0.1 & \textrm{---} & \textrm{---}  & \textrm{---}& \textrm{---}\\
\text{J}2022{+}2854 & \text{B}2020{+}28 & 0.7 $\pm$ 0.2 & 0.8 $\pm$ 0.2 & \textrm{---} & \textrm{---}  & \textrm{---}& \textrm{---}\\
\text{J}2022{+}5154 & \text{B}2021{+}51 & 0.2 $\pm$ 0.1 & 0.2 $\pm$ 0.1 & 0.8 $\pm$ 0.1 & 4 & \textrm{---}& \textrm{---}\\
\text{J}2313{+}4253 & \text{B}2310{+}42 & 0.9 $\pm$ 0.1 & 1.0 $\pm$ 0.1 & 0.9 $\pm$ 0.1 & 2  & \textrm{---}& \textrm{---}\\
\enddata
\tablecomments{Estimated screen distance weighted averages and weighted uncertainties, $D_s$, using both transverse velocity (subscript $V$) and arc curvature (subscript $\eta$). Columns with $N$ represent the number of measurements made for that quantity. Some pulsars have multiple arc curvature screen distance estimations due to multiple arcs being present in their secondary spectra.}
\end{deluxetable*}

\begin{figure*}[htbp]
\centering
\includegraphics[width=0.47\textwidth]{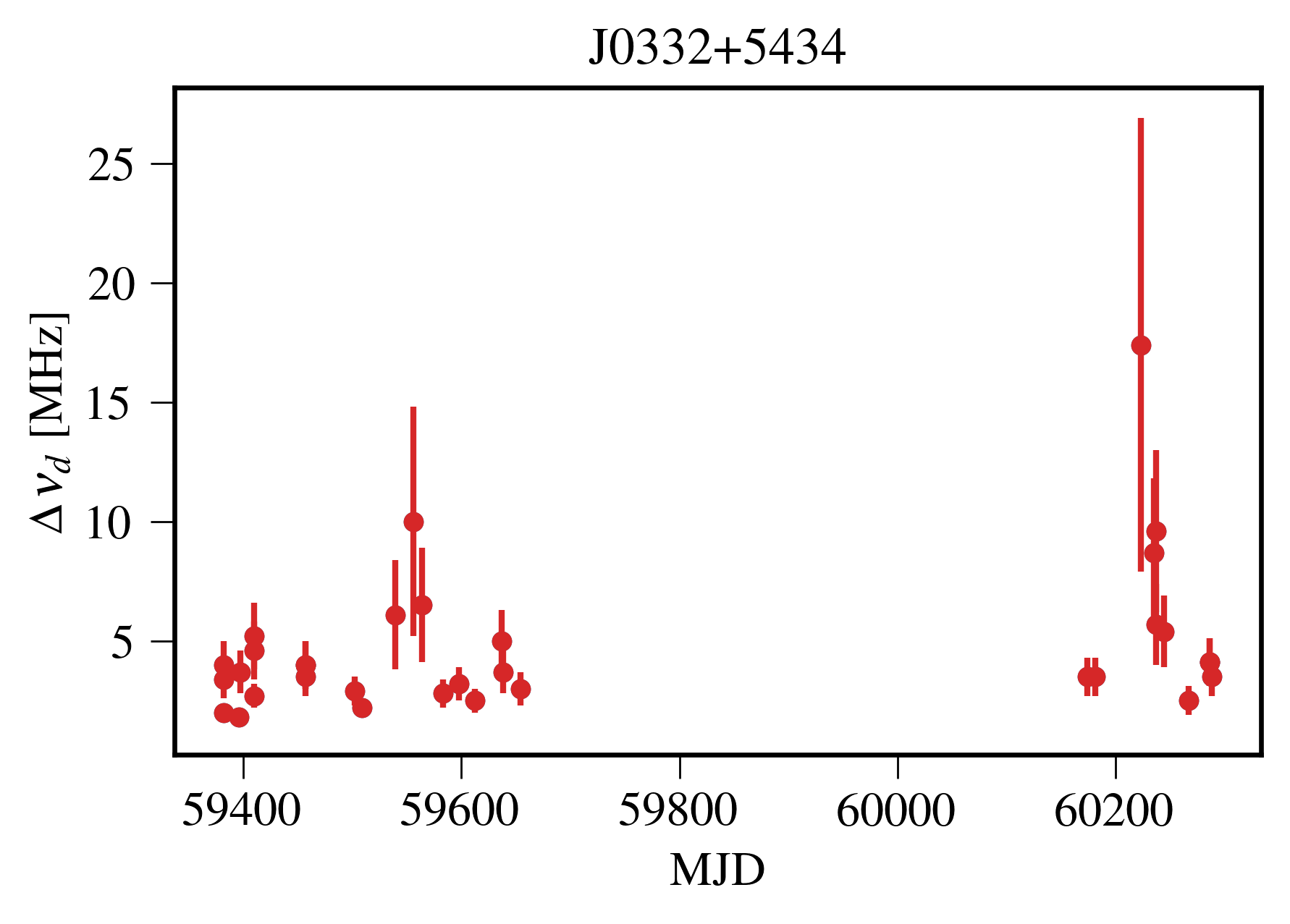}
\hfil
\includegraphics[width=0.47\textwidth]{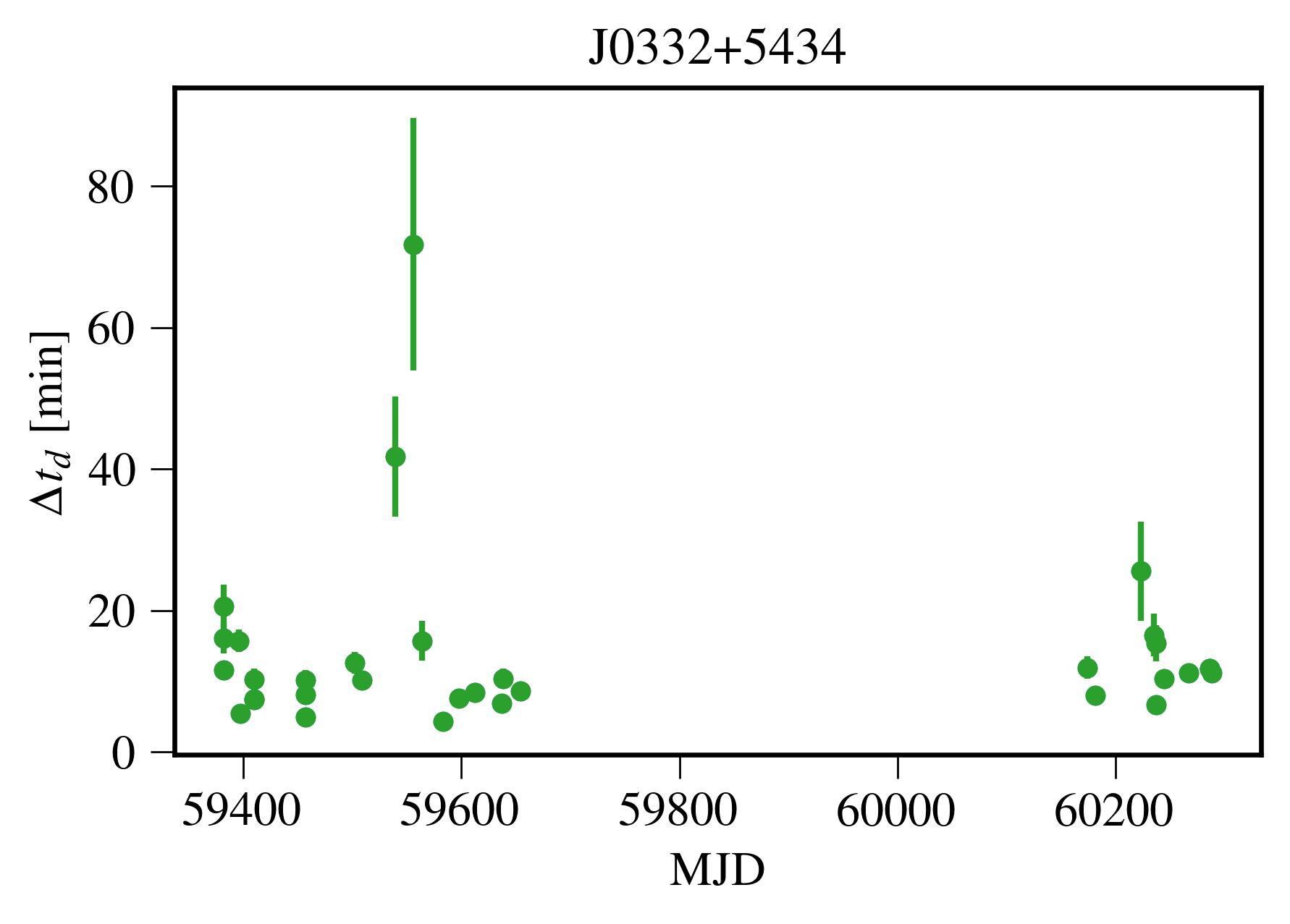}

\medskip
\includegraphics[width=0.47\textwidth]{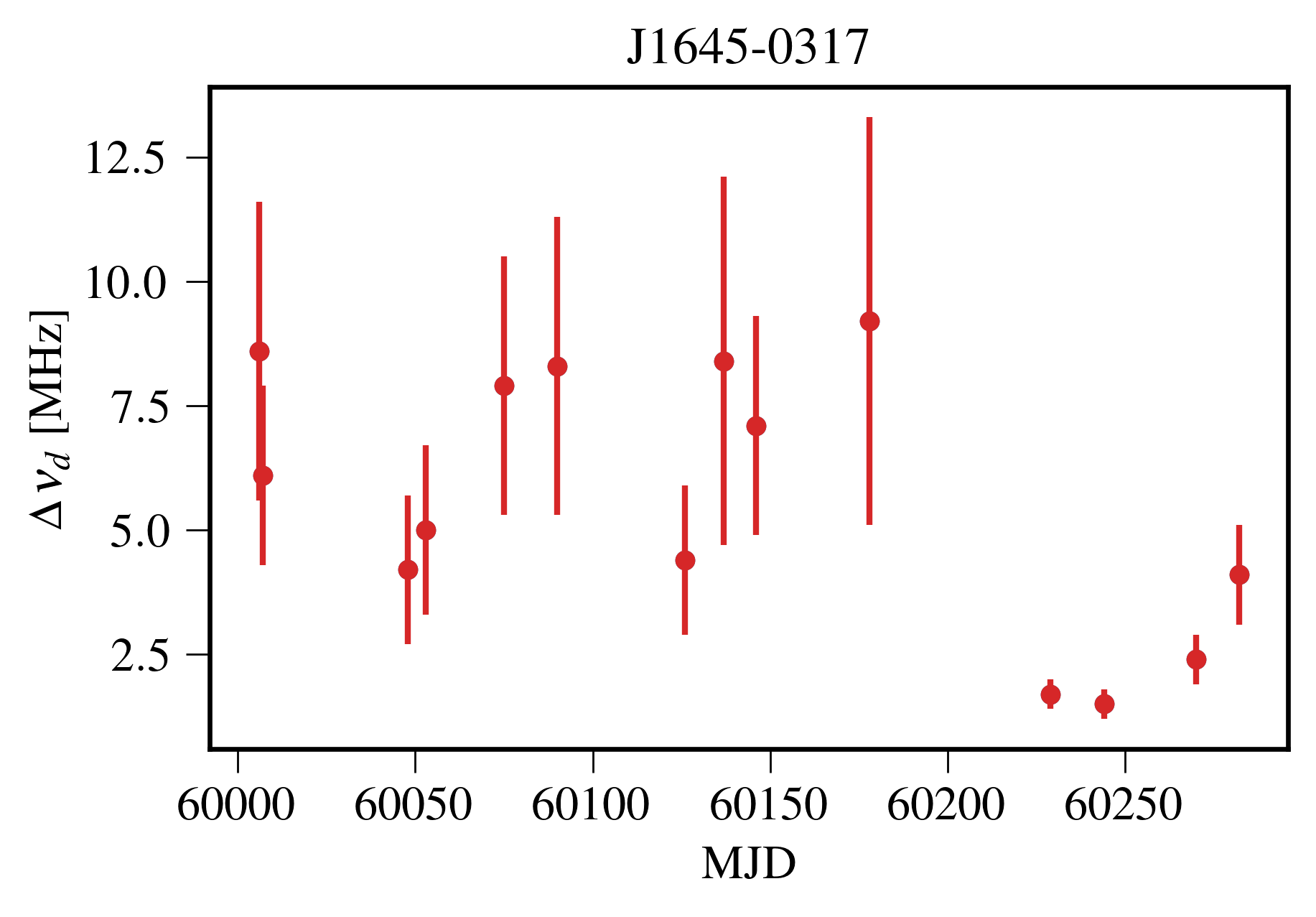}
\hfil
\includegraphics[width=0.47\textwidth]{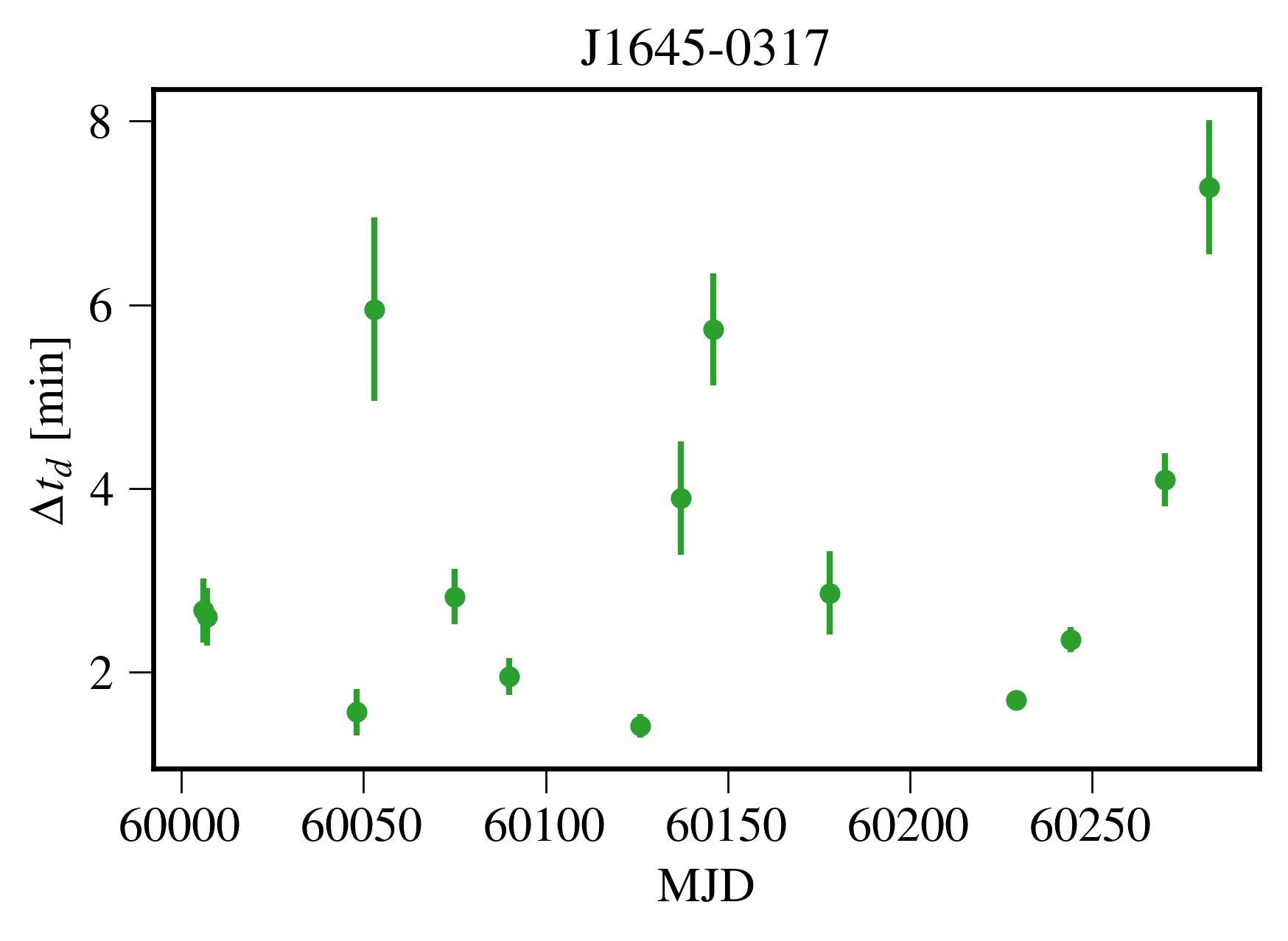}

\medskip
\includegraphics[width=0.47\textwidth]{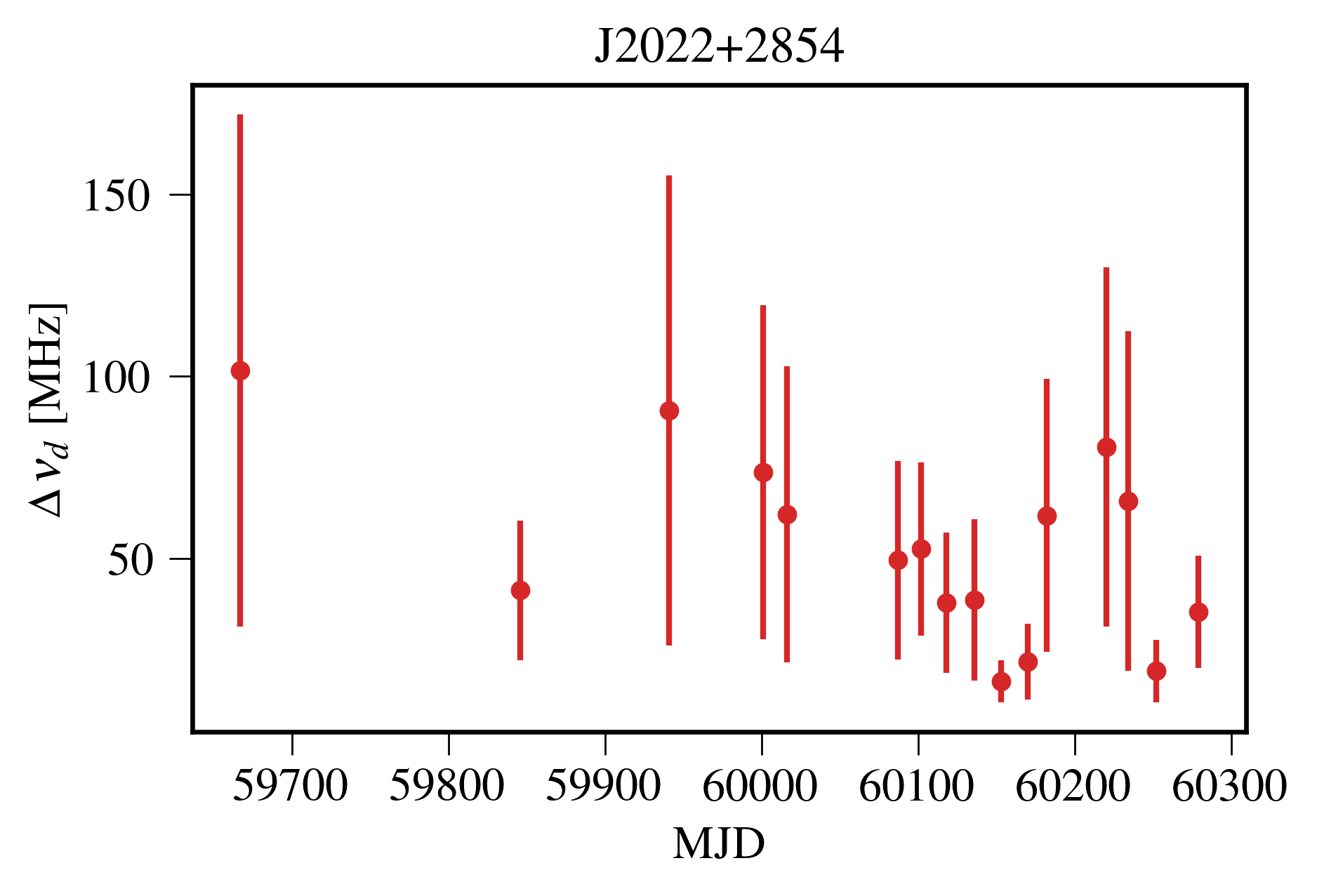}
\hfil
\includegraphics[width=0.47\textwidth]{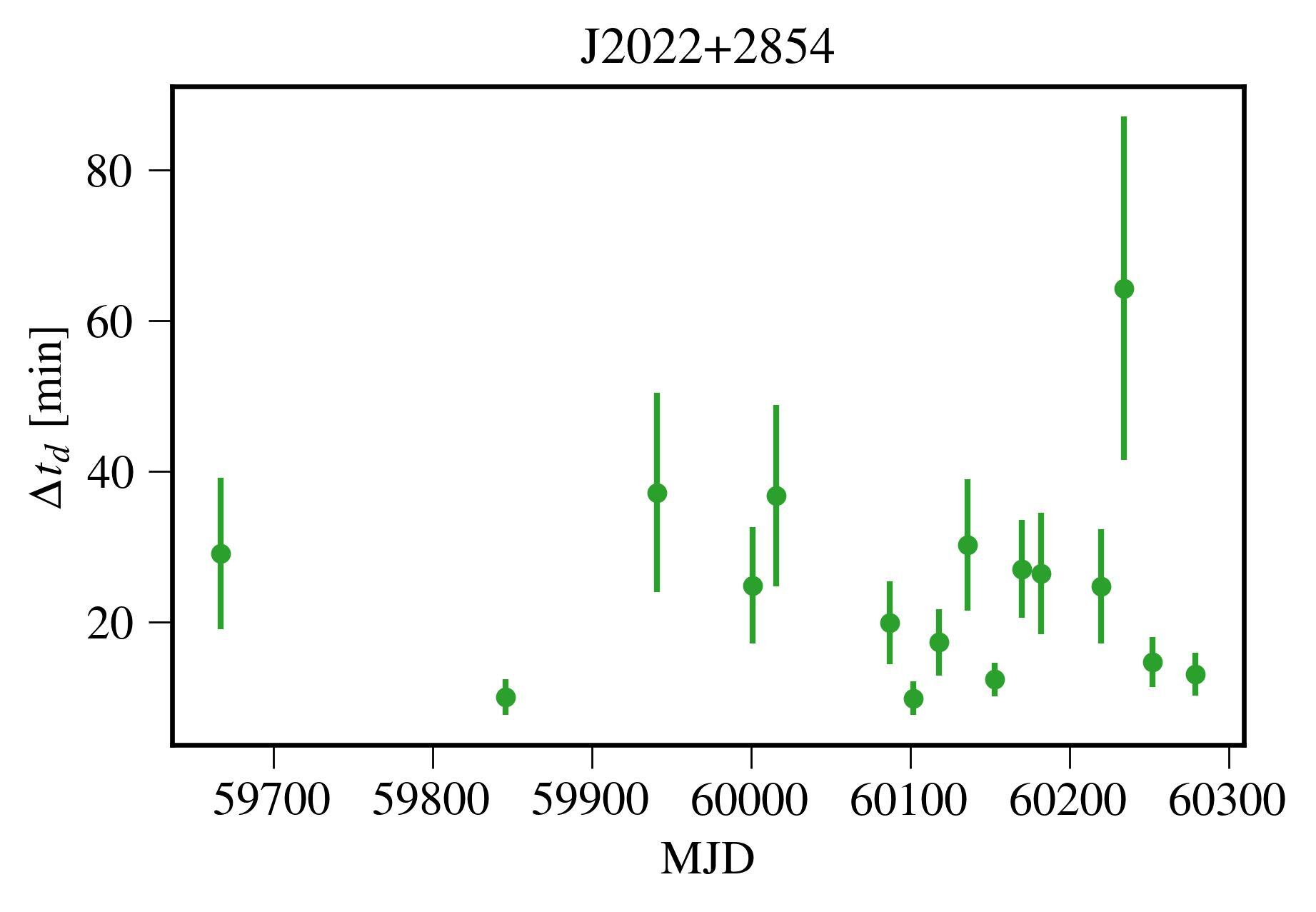}

\medskip
\includegraphics[width=0.47\textwidth]{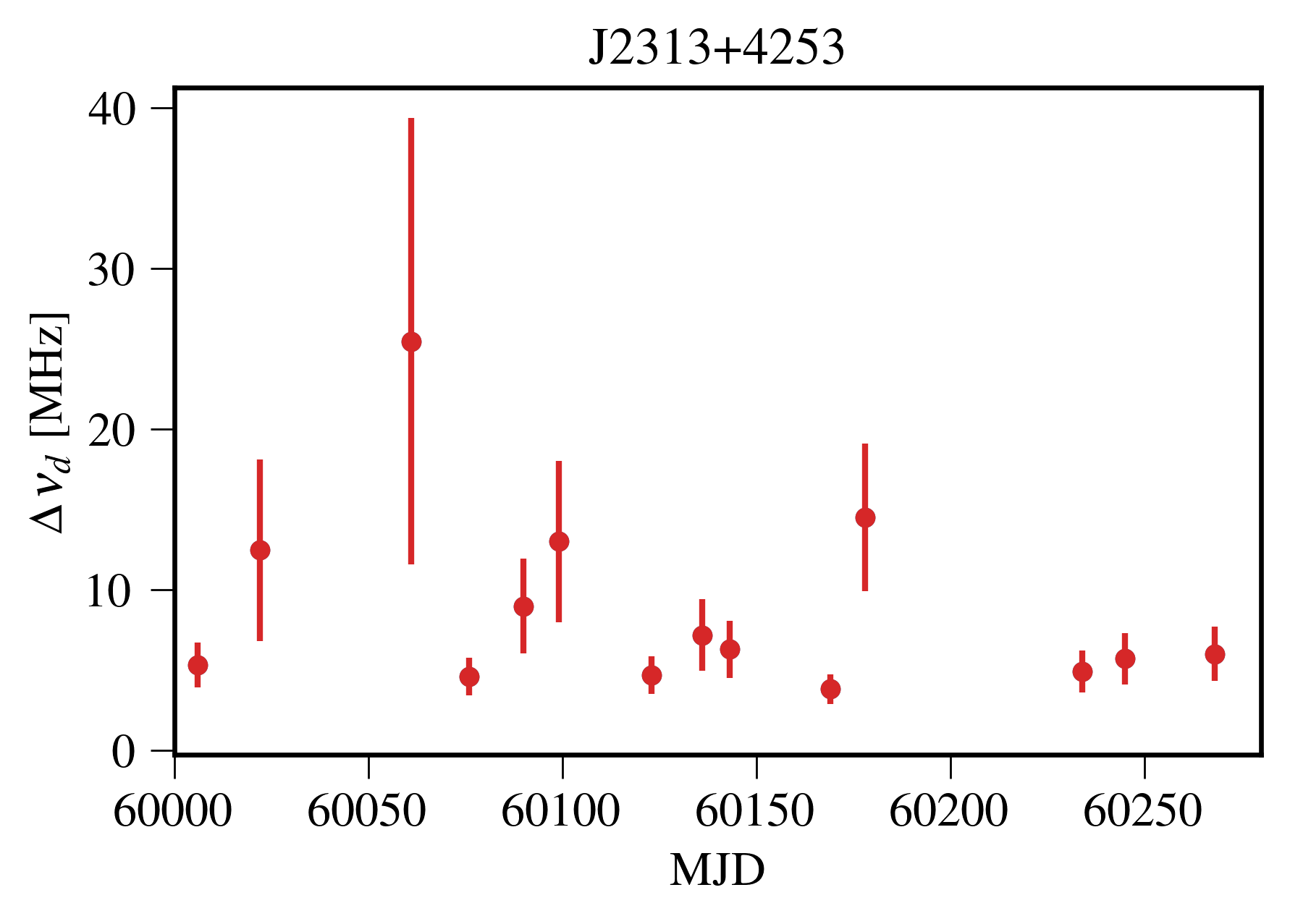}
\hfil
\includegraphics[width=0.47\textwidth]{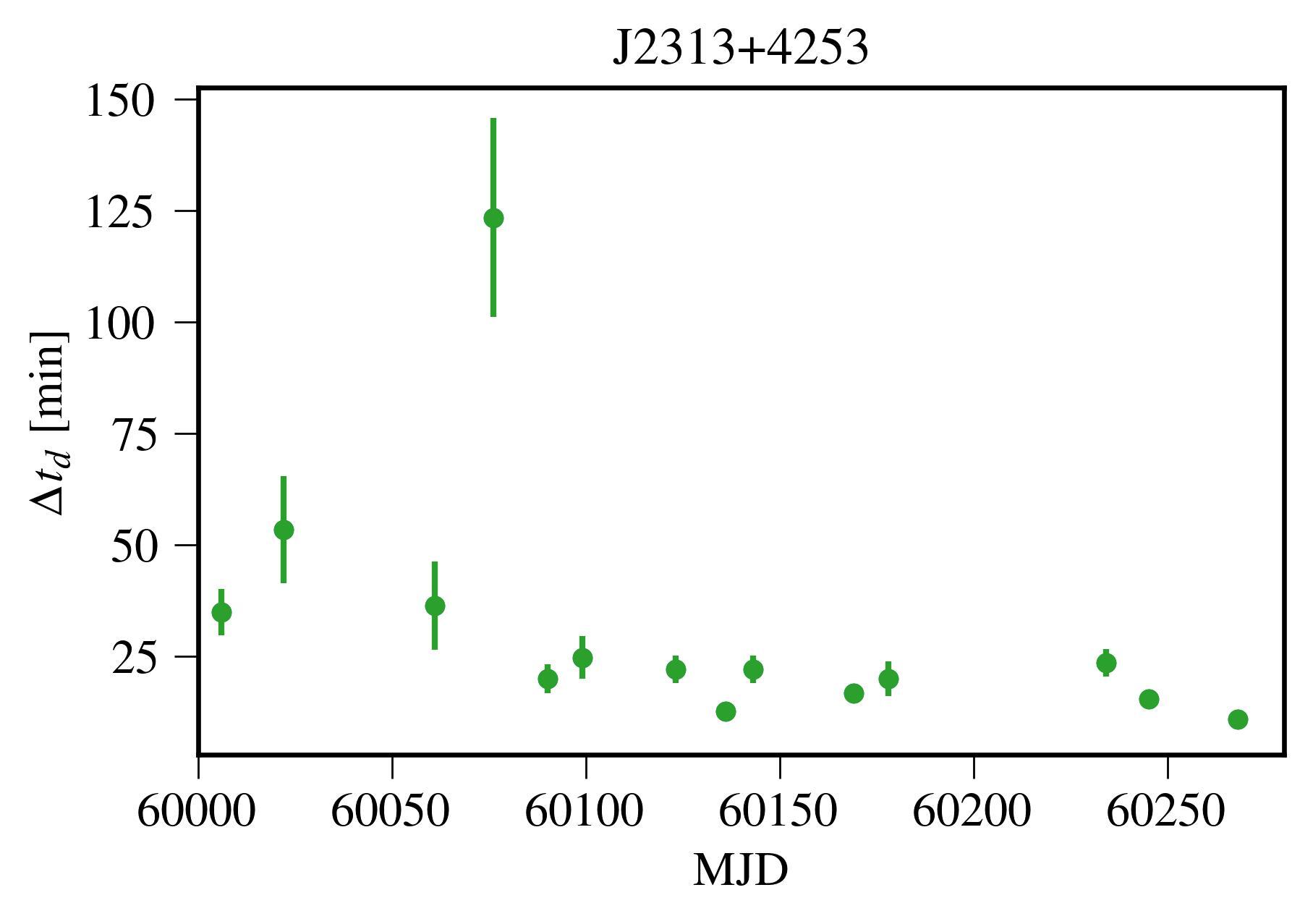}
\caption{Scintillation bandwidth (left) and timescale (right) timeseries for pulsars with at least ten measurements for both quantities.}
\label{timeseries}
\end{figure*}

Our measured scintillation parameters and screen distance estimations, both represented by weighted averages, are shown in Tables \ref{scint_table} and \ref{screen_table}, respectively. Note that all refractive timescale weighted averages for all pulsars using all methods are less than five days. Given that our highest cadence for any pulsar at any given time was one week, we can conclude that all of our ISM measurement made on different days were independent from each other. Timeseries of both scintillation bandwidth and timescale for pulsars with at least ten measurements of both quantities is shown in Figure \ref{timeseries}. 

\subsection{Comparison with Earlier Measurements at Similar Observing Frequencies}

Given that the ISM can change significantly over time along a given LOS, it is important to compare more recent measurements with those taken years or decades prior to monitor for significant variations in scintillation bandwidth and timescale; see Table \ref{table_compare}.
\par While all of the pulsars we have observed have been monitored for decades through various studies, the vast majority of those observations have been at much lower observing frequencies. Consequentially, our observations can offer a somewhat unique view through the ISM to these pulsars. In fact, as far as we can tell from the literature, our observations of J2018$+$2839 may actually constitute this pulsar's first scintillation study at L-band, which is why it is the only pulsar that has been excluded from the aforementioned table. While we could perform a comparison between many of these lower frequency studies by assuming a particular electron density wavenumber spectrum and scaling their scintillation bandwidths and timescales accordingly, this spectrum may vary significantly across different LOSs, and may even change with time along a particular LOS \citep{Levin_Scat, turner_scat, Turner_2024}. Therefore, to avoid introducing a confounding variable into our analyses, we opted only to compare our measurements to literature studies that were made at similar observing frequencies, i.e., we limited our comparison to other studies done at L-band. However, to perform comparisons at the same observing frequencies, we still needed to account for the frequency scaling of scintillation bandwidth and timescale, albeit, over much more comparable frequencies. We therefore assumed a Kolmogorov medium and scaled scintillation bandwidths and timescales to 1400 MHz as $\nu^{4.4}$ and $\nu^{1.2}$, respectively \citep{Romani,Cordes_1998}.
\par While some of the pulsars we observed have fairly consistent scintillation quantities over timespans ranging from a few years to a few decades, particularly PSRs J0332$+$5434, J0826$+$2637, J1645$-$0317, and J2313$+$4253, some others have undergone much more significant changes. As evidenced by a factor-of-two change in the scintillation bandwidth and timescale, emission from PSR J2022+2854 may be experiencing a greater degree of scattering than 20 years ago. However, this may simply be due to a combination of large uncertainties in the case of our scintillation bandwidth measurements, and, in the case of both our scintillation bandwidths and timescales, a few points with very small uncertainties significantly influencing our weighted averages. These lower uncertainty observations all constitute smaller scintillation bandwidth and timescale measurements. This is chiefly a result of smaller scintles meaning more total scintles across the observing band and therefore a smaller finite scintle error, which is the dominant error in our observing setup. We note that many of our measurements from individual epochs are quite consistent with those reported by \cite{wang_old}. These biases may also at least partially explain what we are seeing with PSR J0922$+$0638, which exhibits an order-of-magnitude difference in scintillation bandwidth and timescale weighted averages from measurements taken only a year or two prior to ours \citep{Wang_2024}. That being said, even when examining individual epochs, there is still noticeable disagreement by factors of 50-300\% in some cases.
\par However, none of these possibilities can explain the order-of-magnitude differences we see in PSR J2022$+$5154 with measurements from both two years ago and 20 years ago; even when accounting for biases incurred from weighted averages, all of our scintillation bandwidth measurements are less than 5 MHz and all but one of our scintillation timescale measurements is lower than 9 minutes. There does appear to be a trend of decreasing scintillation bandwidth and timescale when including the other literature measurements, although the sharp decrease from the values in \cite{stine_survey} still feels particularly abrupt. Interestingly, when examining the dynamic spectra from \cite{stine_survey}, the scintillation structure is entirely unrecognizable compared to ours, with the scintles in our measurements being much narrower, individual scintles exhibiting greater symmetry in both time and frequency space, and terminating with sharp, rather than wispy, edges. This is remarkable given that the scintillation structure seen in that paper has seemingly persisted over many years, as evidenced by the striking similarity in the scintillation structure between \cite{wang_old} and \cite{stine_survey}. These sharp differences are explored in more detail in Section \ref{2022_sec}.

\begin{deluxetable*}{CCCC|CCCCC}[ht]

\centering

\tablecolumns{9}

\tablecaption{Comparison with Previously Published Scintillation Parameters \label{table_compare}}
\tablehead{ \multicolumn{2}{C}{} & \multicolumn{2}{C}{\textbf{This work}} & \multicolumn{5}{C}{\textbf{Previously Published Values}} \\ \multicolumn{2}{C}{\textrm{Pulsar}} & 
\multicolumn{1}{C}{$\overline{\Delta \nu}$_{\text{d}}} &
\multicolumn{1}{C}{$\overline{\Delta t}$_{\text{d}}} &
\multicolumn{1}{C}{$\Delta \nu$_{\text{d, scaled}}} &
\multicolumn{1}{C}{$\Delta t$_{\text{d, scaled}}} &
\multicolumn{1}{C}{$\nu_{\textrm{original}}$} &
\multicolumn{1}{C}{\textrm{Year Observed}}&
\multicolumn{1}{C}{\textrm{Reference}} 
\\  \colhead{J2000 Epoch} & \colhead{B1950 Epoch} & \colhead{\text{(MHz)}} & \colhead{\text{(min)}} &  \colhead{\text{(MHz)}} & \colhead{\text{(min)}} & \colhead{\textrm{(MHz)}}& \colhead{} & \colhead{}\vspace{0.15cm}}
\startdata 
\text{J}0332{+}5434 & \text{B}0329{+}54 & 2.8 $\pm$0.7 & 7.6 $\pm$ 1.0 & 2.9 $\pm$ 0.5 & >5.9 & 1420 & 1973 & \textrm{\cite{Wolszczan}} \\
\rotatebox{90}{$\,=$} & \rotatebox{90}{$\,=$} & \rotatebox{90}{$\,=$} & \rotatebox{90}{$\,=$} & 9.2 $\pm$ 0.7 & 15.1 $\pm$ 0.4 & 1540 & 2001-2002 & \textrm{\cite{wang_old}} \\
\rotatebox{90}{$\,=$} & \rotatebox{90}{$\,=$} & \rotatebox{90}{$\,=$} & \rotatebox{90}{$\,=$} & 6.0 $\pm$ 1.4 & 15.3 $\pm$ 2.9 & 1540 & 2004 & \textrm{\cite{wang_new}} \\
\rotatebox{90}{$\,=$} & \rotatebox{90}{$\,=$} & \rotatebox{90}{$\,=$} & \rotatebox{90}{$\,=$} & 5.7 & 15.4 & 1412 & 2011 & \textrm{\cite{Russians}} \\
\text{J}0826{+}2637 & \text{B}0823{+}26 & 21.8 $\pm$ 11.3 & 3.1 $\pm$ 0.5 & 53.9 $\pm$ 3 & 12.1 $\pm$ 0.7 & 1540 & 2001 &  \textrm{\cite{wang_old}} \\
\rotatebox{90}{$\,=$} & \rotatebox{90}{$\,=$} & \rotatebox{90}{$\,=$} & \rotatebox{90}{$\,=$} & 34 $\pm$ 1 & 15.3 $\pm$ 1.3 & 1700 & 2003-2006 &  \textrm{\cite{daszuta}} \\
\text{J}0922{+}0638 & \text{B}0919{+}06 & 4.4 $\pm$ 1.6 & 3.2 $\pm$ 0.3 & 42.6 $\pm$ 12.4 & 16.7 $\pm$ 4.6 & 1250 & 2020-2022 & \textrm{\cite{Wang_2024}} \\
\text{J}1645\text{$--$}0317 & \text{B}1642\text{$--$}03 & 2.1 $\pm$ 0.7 & 2.0 $\pm$ 0.2 & 0.6 & 0.4 & 1412 & 2011 & \textrm{\cite{Russians}}\\
\text{J}2022{+}2854 & \text{B}2020{+}28 & 25.1 $\pm$ 15.1 & 14.1 $\pm$ 4.0 & 46 $\pm$ 3 & 28 $\pm$ 2 & 1540 & 2001-2002 &  \textrm{\cite{wang_old}}\\
\text{J}2022{+}5154 & \text{B}2021{+}51 & 1.1 $\pm$ 0.2 & 3.8 $\pm$ 0.2 & 34 $\pm$ 2 & 32 $\pm$ 3 & 1540 & 2002 & \textrm{\cite{wang_old}}\\
\rotatebox{90}{$\,=$} & \rotatebox{90}{$\,=$} & \rotatebox{90}{$\,=$} & \rotatebox{90}{$\,=$} & 11.5 $\pm$ 17.3 & 20.5 $\pm$ 30.8 & 1400 & 2020 & \textrm{\cite{stine_survey}}\\
\text{J}2313{+}4253 & \text{B}2310{+}42 & 5.2 $\pm$ 1.6 & 16.2 $\pm$ 2.6 & 4.7$\pm$ 5.0 & 17.5 $\pm$ 19.0 & 1400 & 2020 & \textrm{\cite{stine_survey}}
\enddata
\tablecomments{Literature values were reported at observing frequency $\nu_{\textrm{original}}$. Uncertainties represent 1$\sigma$ errors. Scintillation bandwidths were scaled to 1400 MHz assuming an index of 4.4 and scintillation timescales were scaled to 1400 MHz assuming a scaling index of 1.2. We note that this scaling behavior can be inconsistent along different LOSs to different pulsars, and so limit our comparison to observations made at similar frequencies as a precaution.\iffalse \\  $^{*}$Original publication only reported scaled quantities. \\ $^{\dagger}$Values were found within a given range, but no average was reported.\fi}
\end{deluxetable*}
\subsection{Comparison with NE2001}
One of the most commonly used electron density models of the Milky Way is NE2001 \citep{NE2001}, which accounts for Galactic electron density both analytically in addition to considering significant structures local to our region of the galaxy. It can use a pulsar's Galactic position as well as its distance to estimate its dispersion measure, or can conversely use its dispersion measure in conjunction with Galactic position to estimate distance. Additionally, this model can predict scintillation parameters at a given distance along a given LOS through estimation of electron density fluctuations. However, this model, and often other electron density models, depend on measurements that are years, or even decades old. Consequentially, comparisons of modern measurements from pulsars that have been known for decades, such as the ones in our sample, with predictions of these models are useful ways of checking their continued validity. Perhaps more importantly, they are also a valuable way to monitor how the ISM is evolving on long timescales. Additionally, it is important to update these models as additional discrete, highly localized scattering screens are found across various LOSs.
\par \cite{turner_scat} examined how NE2001's predicted scattering delays for 25 of the pulsars in NANOGrav's 12.5-year data set compared with weighted average measurements of these pulsars using the most recent three years of data at the time and found strong agreement with these predictions, achieving a linear correlation coefficient of 0.83 when the data was presented logarithmically. However, as of the writing of our paper, those measurements are anywhere from seven to 11 years old, with the development of NE2001 as old relative to some of those measurements as those measurements are to present day as of this paper's writing. That, along with a large sample of measurements from many pulsars with diverse LOSs, makes a combined comparison of both our data and \cite{turner_scat} with NE2001 a compelling way to examine how the ISM, in some degree of aggregate, may have evolved over the multiple decades since the development of this model. Therefore, to further test the validity of this model against additional LOS, we augmented the results presented in \cite{turner_scat} with our own.

\begin{figure}[!ht]
\hspace*{-0.75cm}
\centering
\includegraphics[scale = 0.6]{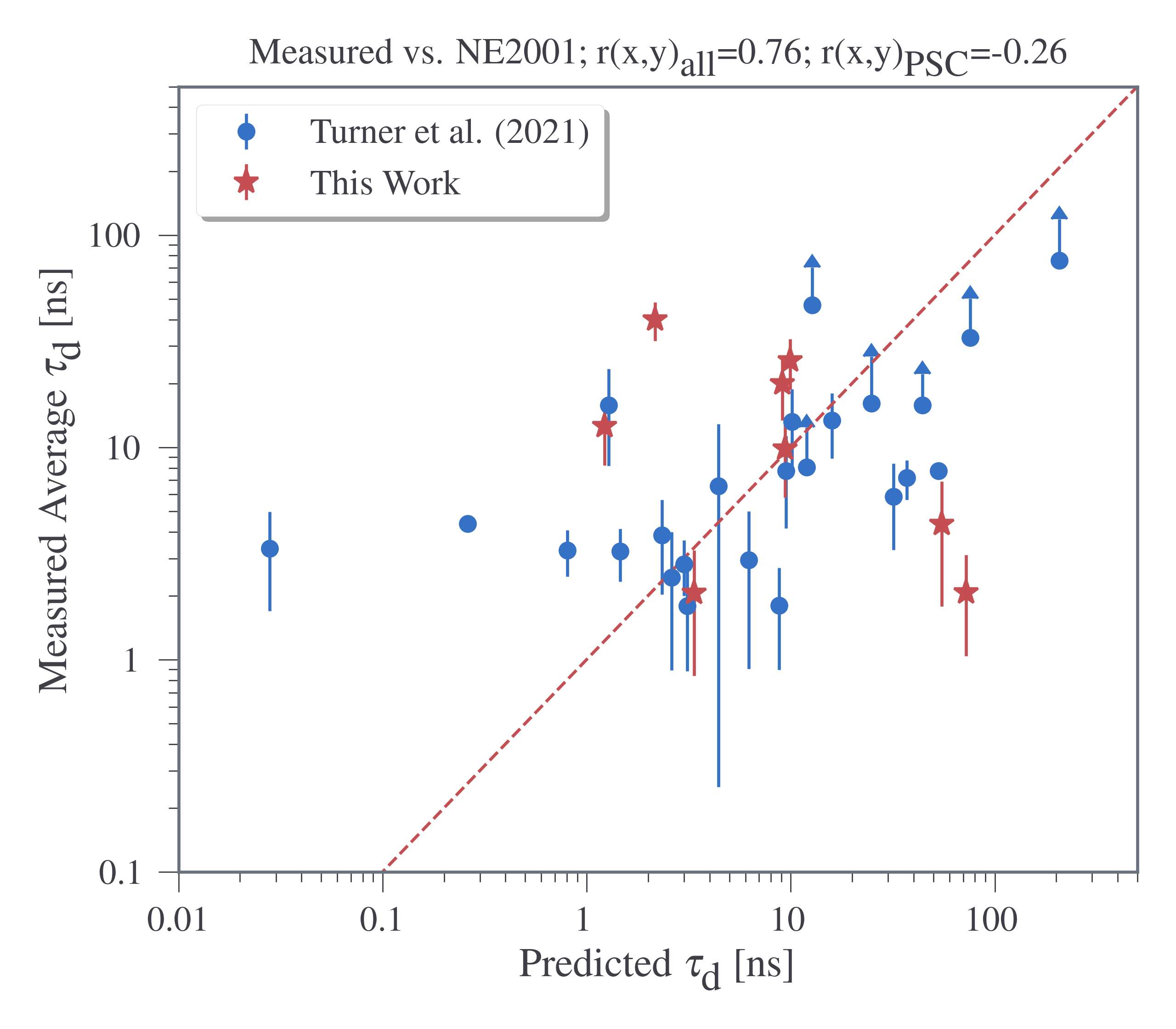}
\caption{Scattering delays at 1500 MHz predicted by NE2001 and measured for pulsars in both \cite{turner_scat} (blue circles) and this work (red stars). Arrows on some of the blue circles indicate lower limits. The dashed red line represents one-to-one correspondence. The high correlation coefficient for the data as a whole indicates general agreement with the model along the LOSs used in these analyses.}
\label{ne2001_compare}
\end{figure}

First, we estimated the scattering delay, $\tau_{\rm d}$ associated with each scintillation bandwidth via

\begin{equation}
    \tau_{\rm d}=\frac{C_1}{2\pi\Delta \nu_{\rm d}},
\end{equation}
where ${C}_{{\rm{1}}}$ is unitless and ranges $0.6-1.5$ depending on the spectrum and geometry of the medium's electron density fluctuations \citep{Cordes_1998}. For consistency with \cite{turner_scat}, we assumed $C_1=1$ in all cases. Since the data in \cite{turner_scat} accomplished this analysis at an observing frequency of 1500 MHz, we scaled our scattering delays to their equivalent values at this frequency. We then combined our data with that of \cite{turner_scat} and determined the resulting weighted linear correlation coefficient in relation to the corresponding predicted scattering delays (determined from the predicted scintillation bandwidths) of NE2001. The results are shown in Figure \ref{ne2001_compare}. After augmentation with our data, the resulting trend and weighted correlation coefficient remain fairly consistent, shifting from 0.91 to 0.76 when only using non-upper limits from \cite{turner_scat}. While this may partially confirm the overall validity of this model, the inclusion of newer data and the resulting decrease in correlation coefficient may possibly indicate a change in the efficacy of this model as a predictor. Relatedly, when examining the correlation coefficient of just our data with NE2001, we end up with a negative correlation, which is surprising, given that the pulsars used in this study have been known for decades and whose scintillation measurements may have been used in the development of NE2001. While this may partially be a selection effect due to our small sample size, as the spread in our data is quite similar to that seen in \citep{turner_scat}, this may also indicate the ISM is evolving significantly enough on the order of one or two decades that updates to this model are required for accurate predictions of newer measurements.

\subsection{Variation of Scattering Delays With Dispersion Measure}
\par Scattering screens can persist along various LOSs for decades \citep{McKee_2022}. Additionally, while the multipath propagation of pulsar emission through these screens can result in consistent structure in observed dynamic spectra over periods of years, changes in the Earth-screen-pulsar LOS over individual epochs can result in slightly varying distributions of scattering angles through a given screen. As a result, measured delays due to interstellar scattering of a pulsar signal can generally be treated as a stationary random process over sufficiently short timescales. Studying this variability over time is crucial to endeavors such as pulsar timing, since monitoring scattering delays on an individual epoch level will become increasingly important as their sensitivity towards potential gravitational wave signals in their data increases. In fact, PTAs are already approaching sensitivities where the most highly scattered pulsars have delays similar to or larger than the uncertainties in their median times-of-arrival \citep{geiger_1903,main_leap,Agazie_2023_timing}, making studies of this effect especially relevant.
\par A possible metric for assessing the sustainability of pulsars for PTAs could involve predicting the expected variability of scattering delays for a given DM. While a relation between scattering delay and dispersion measure has been known for decades \citep{Ramachandran,Bhat_2004,Krishnakumar,Cordes_2016}, it has only relatively recently been demonstrated that the degree of variability in scattering delay is correlated with dispersion measure, indicating a positive relation between total electron density and the fluctuations within that electron density over time \citep{turner_scat}. To further confirm this trend, much like in the previous subsection, we augmented the results of \cite{turner_scat} with our own data. We determined the variability of scattering in each of the pulsars studied by taking the scattering delays inferred in the previous subsection and employing a reduced $\chi^2$ of the form

\begin{equation}
\label{red_chi}
\chi_r^2 = \frac{1}{N -1}\sum_{i=1}^N \frac{(\tau_{\rm d,i}-\overline{\tau_{\rm d}})^2}{\sigma^2_{\tau_{\rm d},i}}.
\end{equation}

\noindent While \cite{turner_scat} calculated these values in two frequency bands (820 MHz and 1500 MHz), for pulsars in which both frequency bands had $\chi_r^2$ values, there was no significant difference in the values determined at both frequencies. As a result, we do not scale our values prior to augmenting their results.

\begin{figure}[!ht]
\centering
\includegraphics[scale = 0.58]{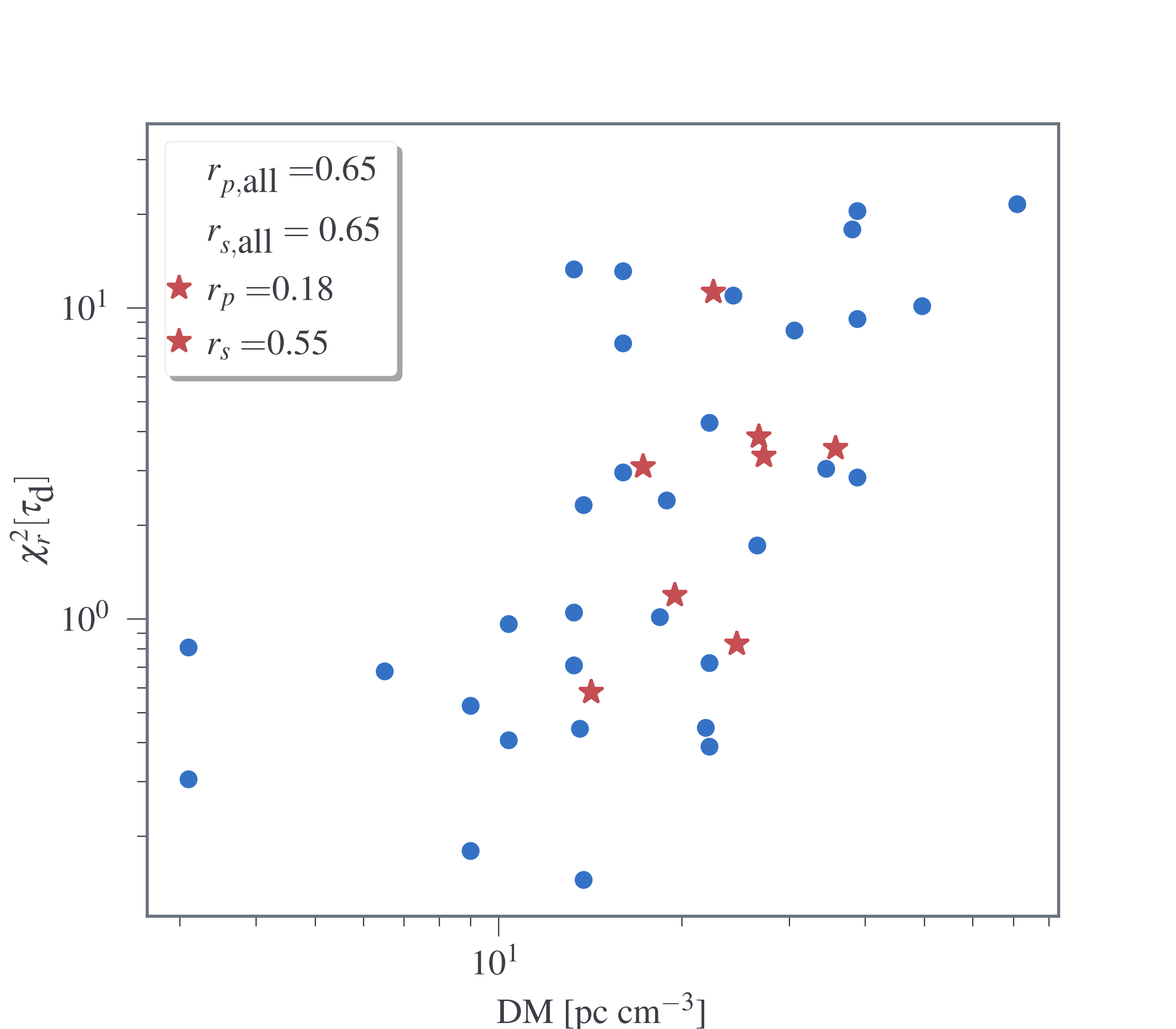}
\caption{Log-log plot of $\chi_r^2$ and the corresponding dispersion measures for pulsars in both \cite{turner_scat} (blue circles) and this work (red stars). The modest Pearson ($r_p$) and Spearman ($r_s$) correlation coefficients  suggests a notable relationship between these two quantities.}
\label{chi_trend}
\end{figure}

As in \cite{turner_scat}, we wanted to account for the possibility of existing correlations being linear or non-linear. For linear correlations, we examined the relation between $\chi_r^2$ and dispersion measure using both the Pearson correlation coefficient,

\begin{equation}
\label{pearson}
r_{p} = \frac{\sigma_{\textrm{DM},\chi_r^2}^2}{\sqrt{\sigma_{\chi_r^2}^2 \sigma_{\textrm{DM}}^2}},
\end{equation} 
where $\sigma_{\textrm{DM},\chi_r^2}^2$ is the covariance between $\textrm{DM}$ and $\chi_r^2$, and $\sigma_{\textrm{DM}}$ and $\sigma_{\chi_r^2}$ are the variances of $\textrm{DM}$ and $\chi_r^2$, respectively. For potentially non-linear correlations, we used the Spearman correlation coefficient,

\begin{equation}
\label{spearman}
r_s = 1- \frac{6\sum_{i=1}^N[\textrm{rg}(\chi_{r,i}^2)-\textrm{rg}(\textrm{DM}_i)]^2}{N(N^2-1)},
\end{equation}
where rg($\chi_{r,i}^2$) and rg($\textrm{DM}_i$) are the ranks of the $i^{\textrm{th}}$ values of $\chi_{r}^2$ and $\textrm{DM}$, respectively, and $N$ is the number of data points being used. The results of these analyses using both our data and that from \cite{turner_scat} can be found in Figure \ref{chi_trend}.

After augmentation with our data, the overall trend remains largely unchanged, with the Pearson correlation coefficient only shifting from 0.69 to 0.65 and the Spearman correlation coefficient holding steady at 0.65. The persistence of the trend after the the addition of our data appears to further confirm the validity of the relation between dispersion measure and epoch-to-epoch variation in scattering delay. Additionally, while the Pearson correlation coefficient on our data is quite low at 0.18, our Spearman correlation coefficient is quite comparable to Spearman correlation coefficient of the data as a whole, at 0.55, highlighting the non-linear nature of this relation. These results are of particular importance for the timing of more highly scattered pulsars, as they demonstrate that simply including an average scattering delay in the timing model of one of these pulsars rather than monitoring scattering delays at the individual epoch will introduce significantly larger time-evolving errors than in less scattered pulsars.

\subsection{Features in Specific Pulsars}
\subsubsection{PSR J0332+5434}
\begin{figure}[!ht]
\centering
\hspace*{-.6cm}
\includegraphics[scale = 0.55]{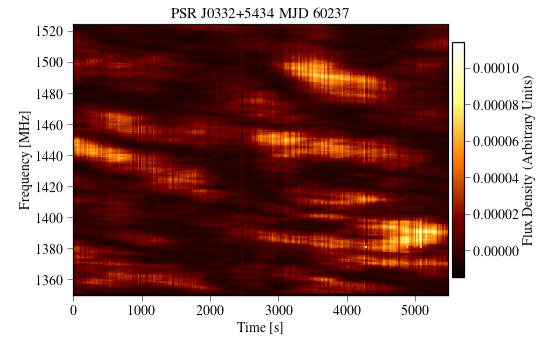}
\caption{Example dynamic spectrum for PSR J0332+5434 on MJD 60237.}
\label{0332_dyn_ex}
\end{figure}

\begin{figure}[!ht]
\centering
\includegraphics[scale = 0.55]{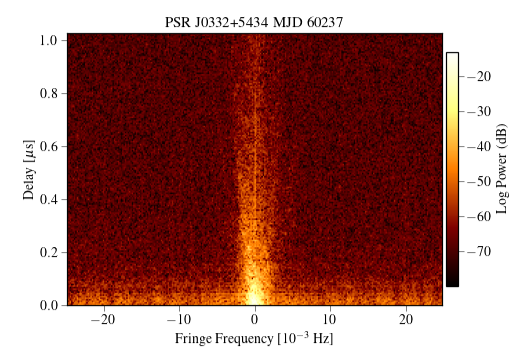}
\caption{Example secondary spectrum for PSR J0332+5434 on MJD 60237.}
\label{0332_sec_ex}
\end{figure}

An example dynamic and secondary spectrum can be seen in Figures \ref{0332_dyn_ex} and \ref{0332_sec_ex}, respectively. This pulsar has least at four known scintillation arcs, all of which seem to be closer to the pulsar than to Earth \citep{Margaret_L_Putney_2006}. In our data, we find evidence for two distinct arcs, which combine with our transverse velocity measurements to yield detections of three difference scattering screens. Although we do not have access to all information used to determine the fractional screen distances in \cite{Margaret_L_Putney_2006}, extrapolating using an updated  distance of 1.68 kpc determined from very long baseline interferometry (VLBI) \citep{Deller_2019}, the data we do have strongly suggests that two of the three screens we detected were also found by \cite{Margaret_L_Putney_2006}. However, the screen we call $D_{s,\eta,1}$ is too close to Earth to correspond to any of the four screens they found, especially after accounting for the updated further distance to the pulsar. We therefore suggest this pulsar may have least five scattering screens, while cautioning that degeneracies in more advanced screen-distance-estimation methodologies may place screens at a similar distance from the pulsar that other potential solutions may place them from Earth. 
\par One of the arcs we observe is quite diffuse and extremely asymmetric, with only one arm usually visible at a time, and the visible arm shifting from the left to the right arm over the course of our observations. This asymmetry could indicate a refractive wedge along the LOS to this pulsar \citep{cordes_2006_refraction}.
\subsubsection{PSR J0826+2637}
\begin{figure}[!ht]
\centering
\hspace*{-.85cm}
\includegraphics[scale = 0.58]{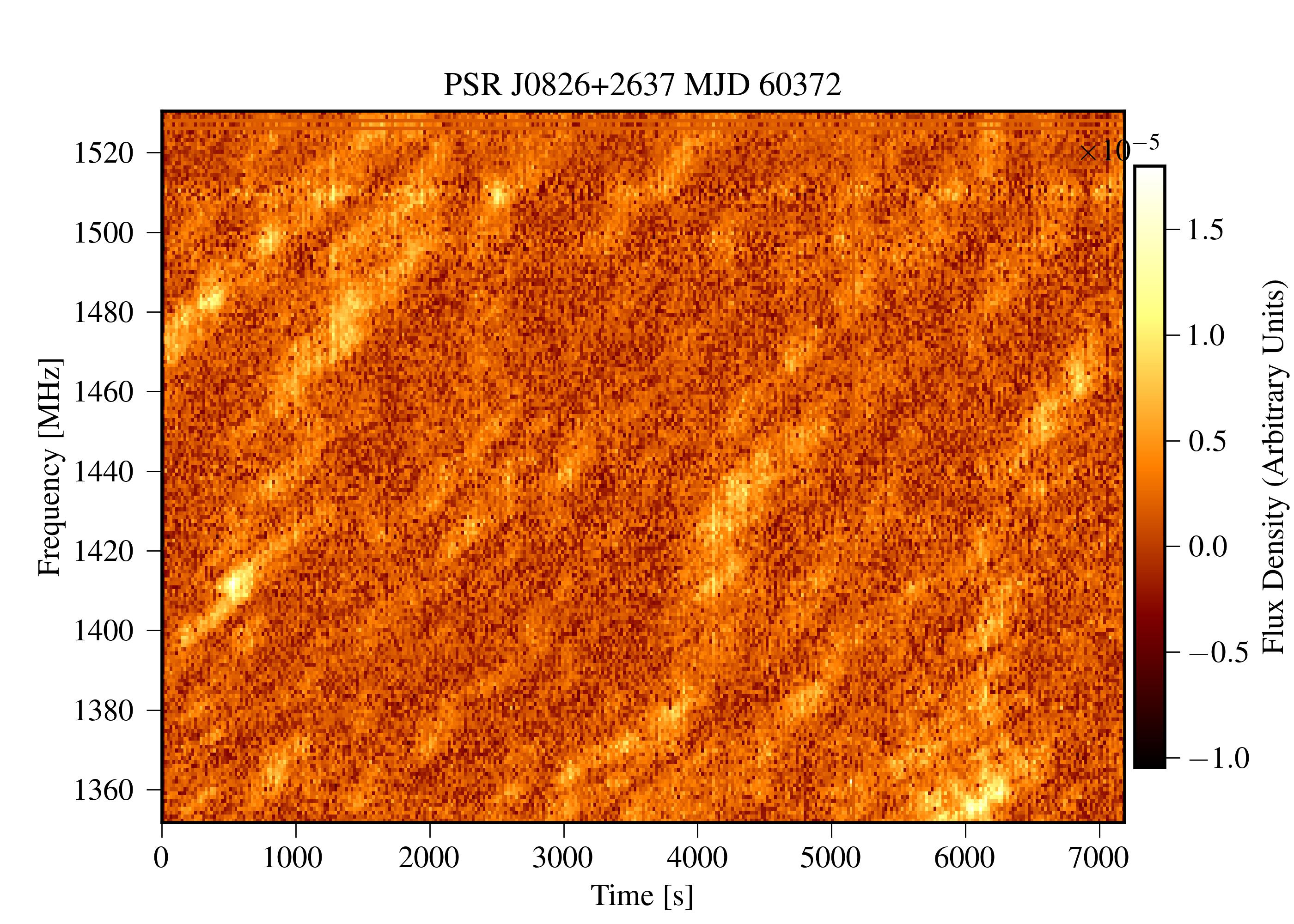}
\caption{Example dynamic spectrum for PSR J0826+2637 on MJD 30372.}
\label{0826_dyn_ex}
\end{figure}

\begin{figure}[!ht]
\centering
\hspace*{-0.85cm}
\includegraphics[scale = 0.105]{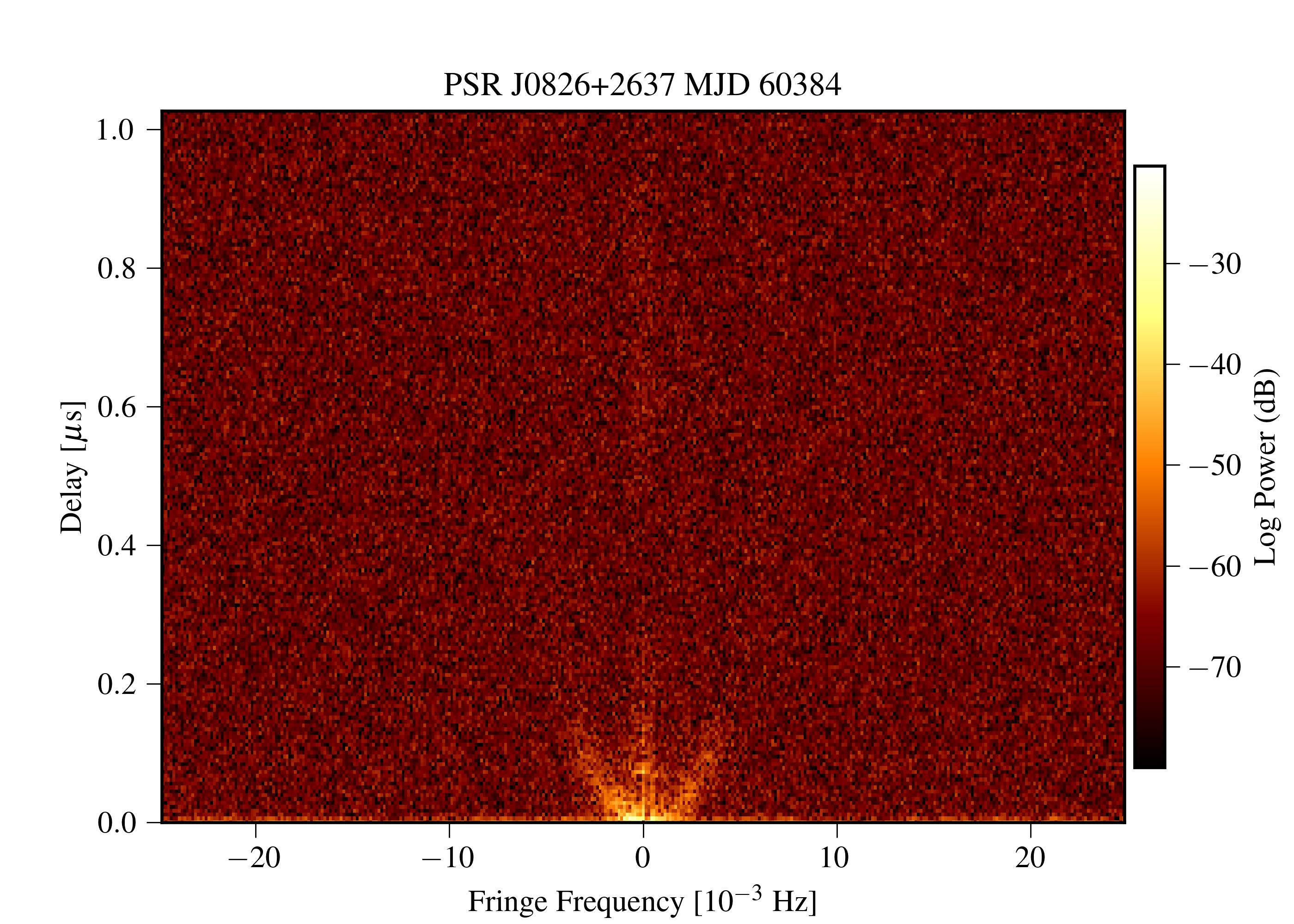}
\caption{Example secondary spectrum for PSR J0826+2637 on MJD 60384.}
\label{0826_sec_ex}
\end{figure}

An example dynamic and secondary spectrum can be seen in Figures \ref{0826_dyn_ex} and \ref{0826_sec_ex}, respectively. Scattering screen distance estimations obtained both from scintillation-derived transverse velocities and from scintillation arcs yield a screen distance that agrees well with the third-furthest of the four screens found by \cite{Margaret_L_Putney_2006}, once factoring in updated pulsar distance measurements \citep{Deller_2019}. The visible arc in this pulsar appears to undergo rapid changes in the power along the arms, with the majority of power going from the right arm, to roughly symmetric between the arms, to the majority of power in the left arm in less than two months. Some of the rapid shift in arm power may be a consequence of refraction, given that we find scintillation drift rate-derived refractive angles around 0.04 mas.
\subsubsection{PSR J0922+0638}
\begin{figure}[!ht]
\centering
\hspace*{-.1cm}
\includegraphics[scale = 0.43]{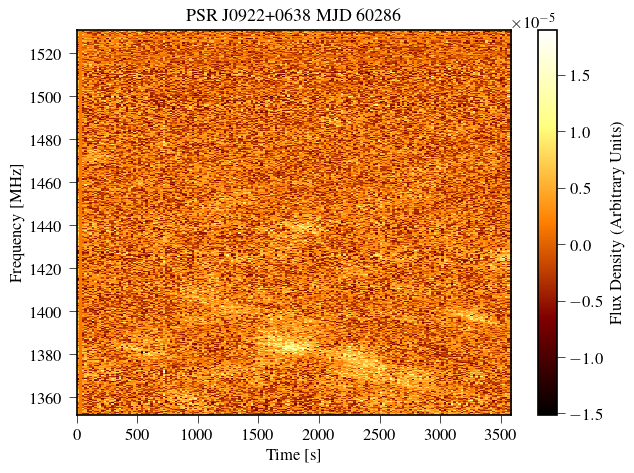}
\caption{Example dynamic spectrum for PSR J0922+0638 on MJD 60286.}
\label{0922_dyn_ex}
\end{figure}

An example dynamic spectrum can be seen in Figure \ref{0922_dyn_ex}. Based on a $V_{\rm ISS}$ analysis, \cite{chatterjee_2001} estimate a scintillation-velocity derived screen around 0.25 kpc from Earth. This is in sharp contrast to the screen distance of $0.9\pm0.1$ kpc from Earth that we find. However, once scaling their highest observing frequency scintillation measurements to our observing frequency assuming a Kolmogorov wavenumber spectrum ($\Delta \nu_{\rm d} \propto \nu^{4.4}; \Delta t_{\rm d} \propto \nu^{1.2}$), we find our scintillation bandwidths are around 45\% larger and our scintillation timescales are around 60\% smaller that what they would have reported at an equivalent observing frequency. This suggests the ISM along the LOS to this pulsar has changed noticeably in the nearly three decades between these sets of observations. This claim, as well as our screen distance estimation, are further justified when comparing our results with those of \cite{ocker_arcs}, who used scintillation arc measurements taken within a year of our observations to conclude possible screen distances ranging from around 0.6$-$1.1 kpc.
\subsubsection{PSR J1645$-$0317}
\begin{figure}[!ht]
\centering
\hspace*{-.5cm}
\includegraphics[scale = 0.55]{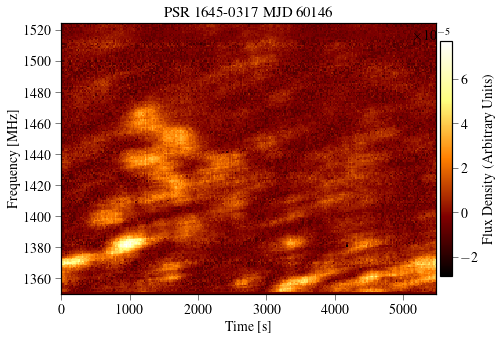}
\caption{Example dynamic spectrum for PSR J1645-0317 on MJD 60146.}
\label{1645_dyn_ex}
\end{figure}

\begin{figure}[!ht]
\centering
\hspace*{-0.8cm}
\includegraphics[scale = 0.56]{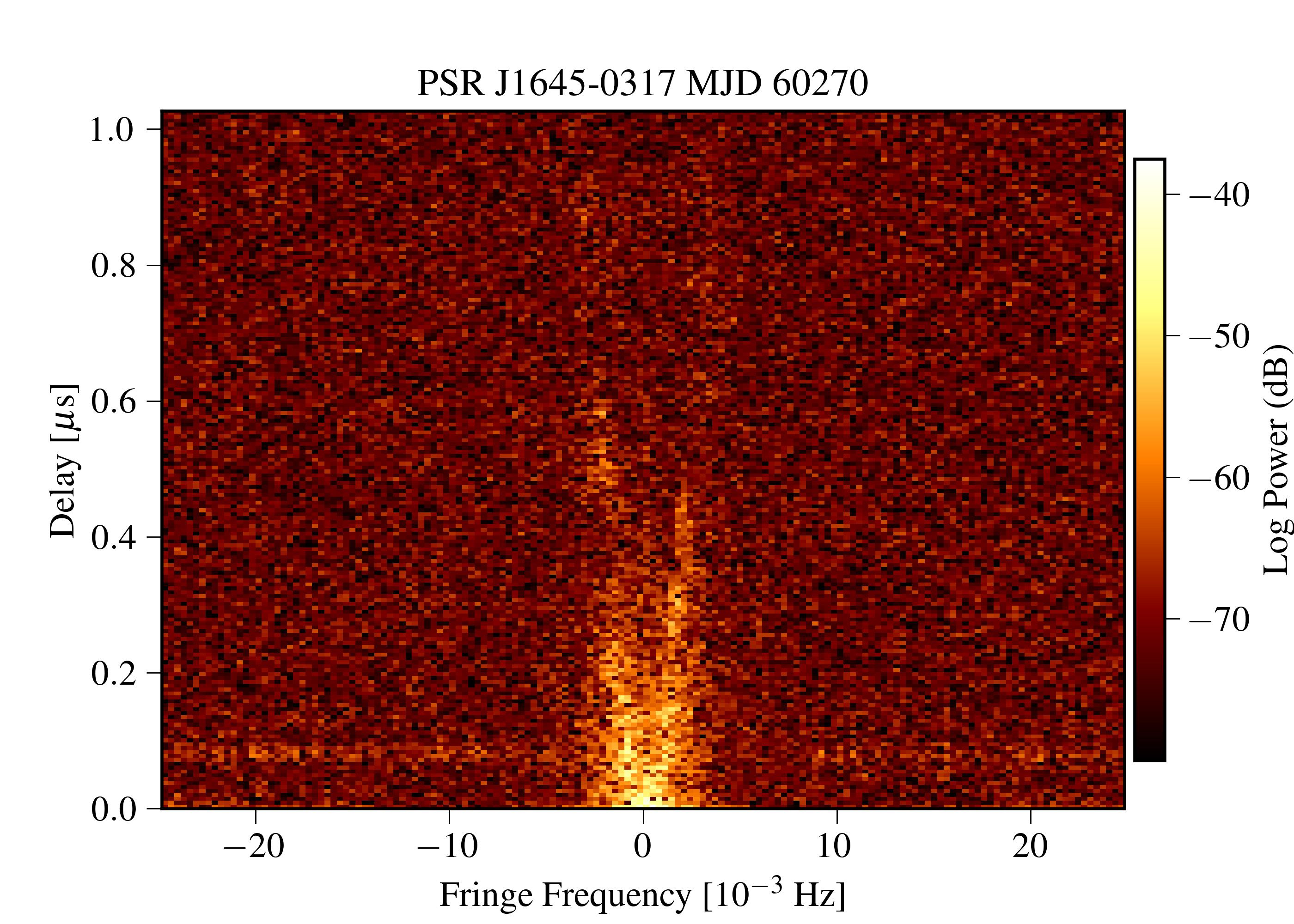}
\caption{Example secondary spectrum for PSR J1645-0317 on MJD 60270.}
\label{1645_sec_ex}
\end{figure}

An example dynamic and secondary spectrum can be seen in Figures \ref{1645_dyn_ex} and \ref{1645_sec_ex}, respectively. The shallowest arc we detect exhibits an asymmetry that shifts from the right arm to the left and then back over timescales of a few months, potentially indicating a refractive wedge along the LOS to this pulsar \citep{cordes_2006_refraction}. Some of the arcs also exhibit a patchiness which may indicated the presence of arclets, which could originate from effects including astronomical-unit scale inhomogeneities within the associated scattering screen \citep{hsa+05} or from a double-lensing effect caused by material closer to the pulsar relative to the screen \citep{Zhu_2023}.
\par In total, this pulsar has three known scintillation arcs, whose fractional screen distances have been estimated in previous works \citep{Margaret_L_Putney_2006}. In this work, we managed to observe two distinct arcs over the course of our observations, although combined with our transverse velocity analyses we detect three distinct scattering screens overall. It is important to note that the fractional screen distance estimates made in \citep{Margaret_L_Putney_2006} used an older estimate of the pulsar's distance, placing it at 2.91 kpc from Earth, whereas we use a more updated measurement achieved via VLBI, placing it at around 3.97 kpc from Earth \citep{Deller_2019}. 
\par
\begin{figure}[!ht]
\centering
\includegraphics[scale = 0.75]{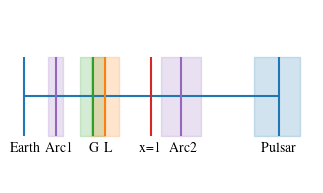}
\caption{Diagram showing screens relative to their distance between Earth and the pulsar PSR J1645$-$0317. The labels ``G" and ``L" represent screens derived from transverse velocities using Gaussian and Lorentzian fits to ACFs, respectively. A scenario in which $x=1$ represents the halfway point between Earth and the pulsar, as defined in Equation \ref{iss_velo}. Lighter shaded bars indicate uncertainties on the associated distances.}
\label{screen_dist}
\end{figure}

 Two of the three arcs found in \cite{Margaret_L_Putney_2006} place scattering screens within around 100 pc of the pulsar, whereas our screen distances are all in the middle third of the distance between Earth and the pulsar. As with PSR J0332+5434, while we do not have access to all information used to determine the fractional screen distances in \cite{Margaret_L_Putney_2006}, the data we do have strongly suggests two of the three screens detected in this work were also found by \cite{Margaret_L_Putney_2006}. Taken together, these results seem to indicate one of our measured scintillation arc scattering screens may be unique to this work, and may indicate this pulsar has at least five scattering screens along its LOS. However, we again note the same degeneracy qualification mentioned in the section on PSR J0332+5434. The screens visible in this work can be seen in Figure \ref{screen_dist}.
\subsubsection{PSR J2018+2839}
\begin{figure}[!ht]
\centering
\hspace*{-.7cm}
\includegraphics[scale = 0.3]{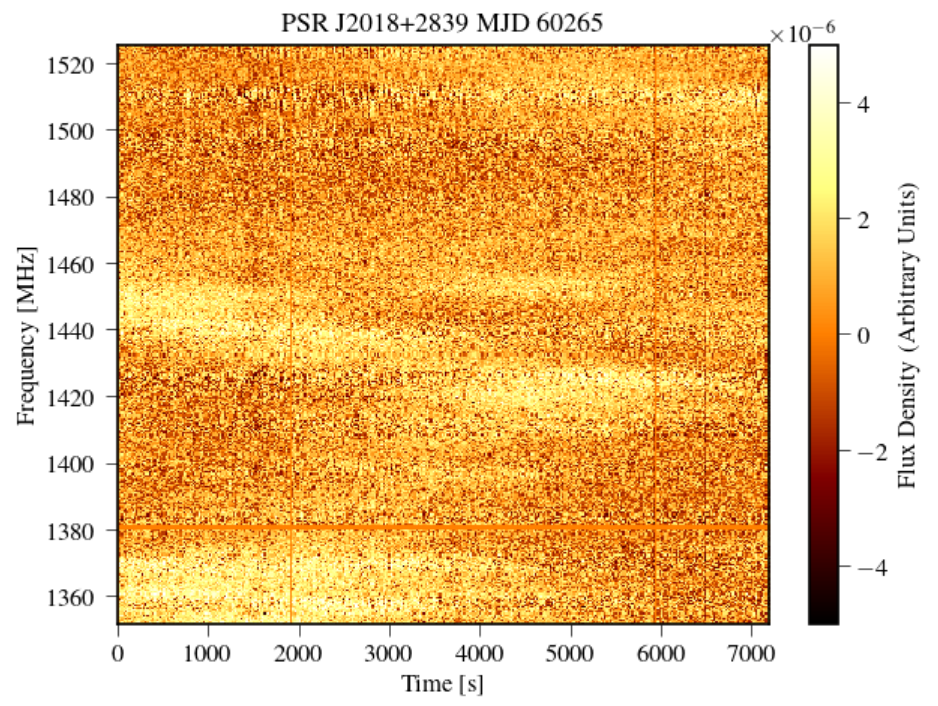}
\caption{Example dynamic spectrum for PSR J2018+2839 on MJD 60265.}
\label{2018_dyn_ex}
\end{figure}

An example dynamic spectrum can be seen in Figure \ref{2018_dyn_ex}. This pulsar had by far the longest weighted average scintillation timescales of any pulsar in our study, ranging from around half an hour to an hour depending on the the type of fit used. While, as mentioned earlier, our observations of this pulsar may constitute its first scintillation study at L-band, this pulsar's LOS appears to be fairly stable even on timescales, exceeding 20 years when accounting for studies done at other observing frequencies. A scatter broadening measurement taken by \cite{kuzmin}, once converting to scintillation bandwidths and scaling up to 1400 MHz under the assumption of a Kolmogorov wavenumber spectrum, yields a scintillation bandwidth of around 9.9 MHz, agreeing with our measurements within error.
\subsubsection{PSR J2022+2854}
\begin{figure}[!ht]
\centering
\hspace*{-.8cm}
\includegraphics[scale = 0.413]{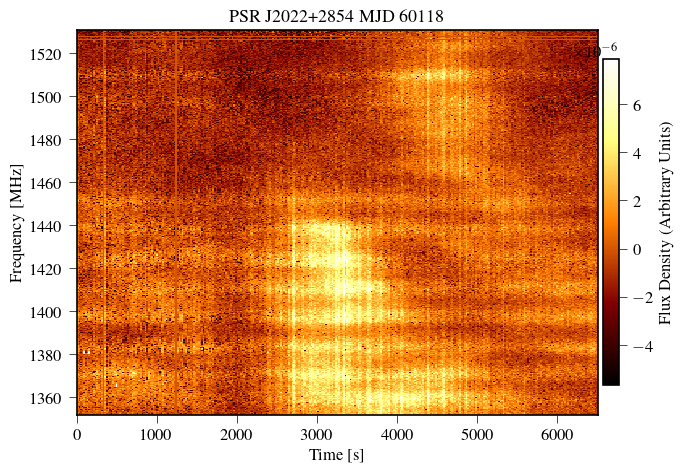}
\caption{Example dynamic spectrum for PSR J2022+2854 on MJD 60265.}
\label{202228_dyn_ex}
\end{figure}

An example dynamic spectrum can be seen in Figure \ref{202228_dyn_ex}. As mentioned earlier, while there may be some evidence of a change in the LOS over the past two decades, once accounting for uncertainties in some measurements biasing our weighted averages and comparing individual values, it is likely that scintillation along this LOS is fairly stable over these timescales. A scatter broadening measurement taken by \cite{kuzmin}, once converting to scintillation bandwidths and scaling up to 1400 MHz under the assumption of a Kolmogorov wavenumber spectrum, yields a scintillation bandwidth of around 23 MHz, agreeing quite well with our measurements within error.
\subsubsection{PSR J2022+5154}
\label{2022_sec}
\begin{figure}[!ht]
\centering
\hspace*{-0.5cm}
\includegraphics[scale = 0.42]{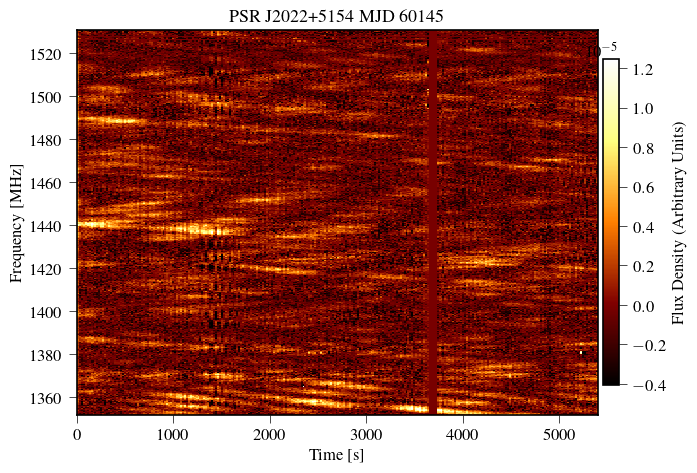}
\caption{Example dynamic spectrum for PSR J2022+5154 on MJD 60145.}
\label{202251_dyn_ex}
\end{figure}

\begin{figure}[!ht]
\centering
\hspace*{-0.5cm}
\includegraphics[scale = 0.58]{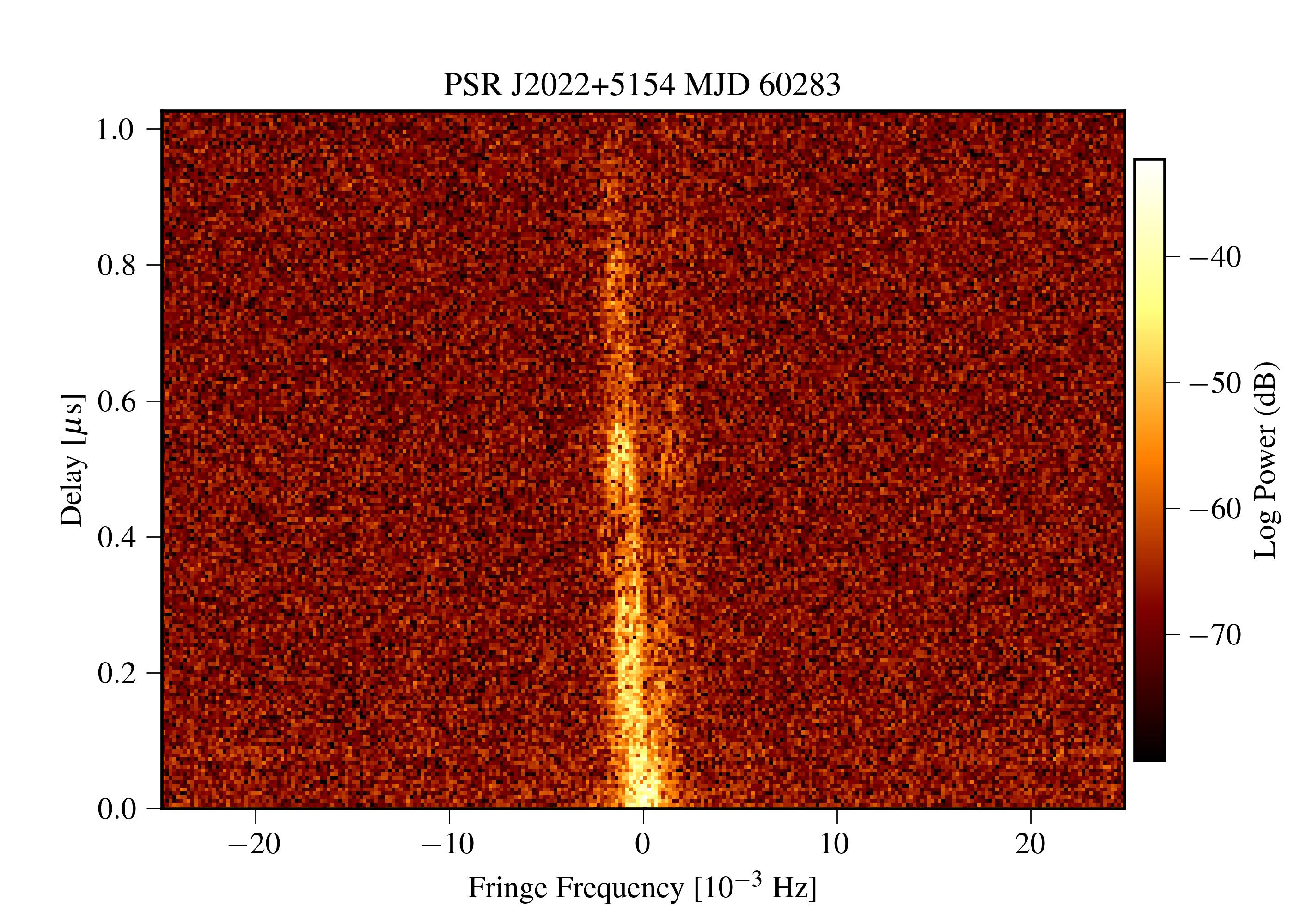}
\caption{Example secondary spectrum for PSR J2022+5154 on MJD 60283.}
\label{202251_sec_ex}
\end{figure}

An example dynamic and secondary spectrum can be seen in Figures \ref{202251_dyn_ex} and \ref{202251_sec_ex}, respectively. The scintillation arc observed in this pulsar exhibits asymmetry that may be periodic with Earth's orbit around the sun. In our first detected arc observation, the power in both arms is roughly symmetric, although the left arm does appear a bit brighter. Our second detected arc 138 days later has much more power in the left arm, although this dominance shifts to the right arm in our observation 23 days later, also remaining consistent in the next detection 22 days afterwards. These shifts in power along the left and right arms may be due to a refractive wedge along to LOS to this pulsar, with our slightly different view through the wedge as the Earth orbits the sun affecting which arm contains more power \citep{cordes_2006_refraction}. 
\par Although our arc curvature measurements match well with those made by \cite{stine_survey} at 1400 MHz, as mentioned earlier, our dynamic spectra look completely different, with both our measured scintillation bandwidths and timescales being an order of magnitude smaller than theirs. While it is not entirely unexpected that our dynamic spectra and scintillation bandwidths are different, as our measurements and theirs constrain this pulsar's refractive timescale to within around 2$-$4 days, and our observations were taken more than three years apart, an order-of-magnitude difference in both quantities is a bit surprising while maintaining the same arc curvature. However, we also note that \cite{stine_survey} saw two scintillation arcs, and therefore two scattering screens, in their data, whereas we only see one of these arcs our in our observations. Taken together, this seems strongly indicative of the single scattering screen we observe dominating the scintillation pattern in our data, and the other of the two screens observed in \cite{stine_survey} dominating the scintillation pattern in their data. This makes the behavior along this LOS somewhat analogous to the double-screen scenario in PSR B1508+55 found in \cite{sprenger}. Additionally, unlike the arcs observed in \cite{stine_survey}, the arcs we observed also exhibit considerable asymmetry, substructure, and thickness, potentially indicating non-uniform scattering in at least one of the observed screens.

\subsubsection{PSR J2313+4253}
\begin{figure}[!ht]
\centering
\hspace*{-0.8cm}
\includegraphics[scale = 0.58]{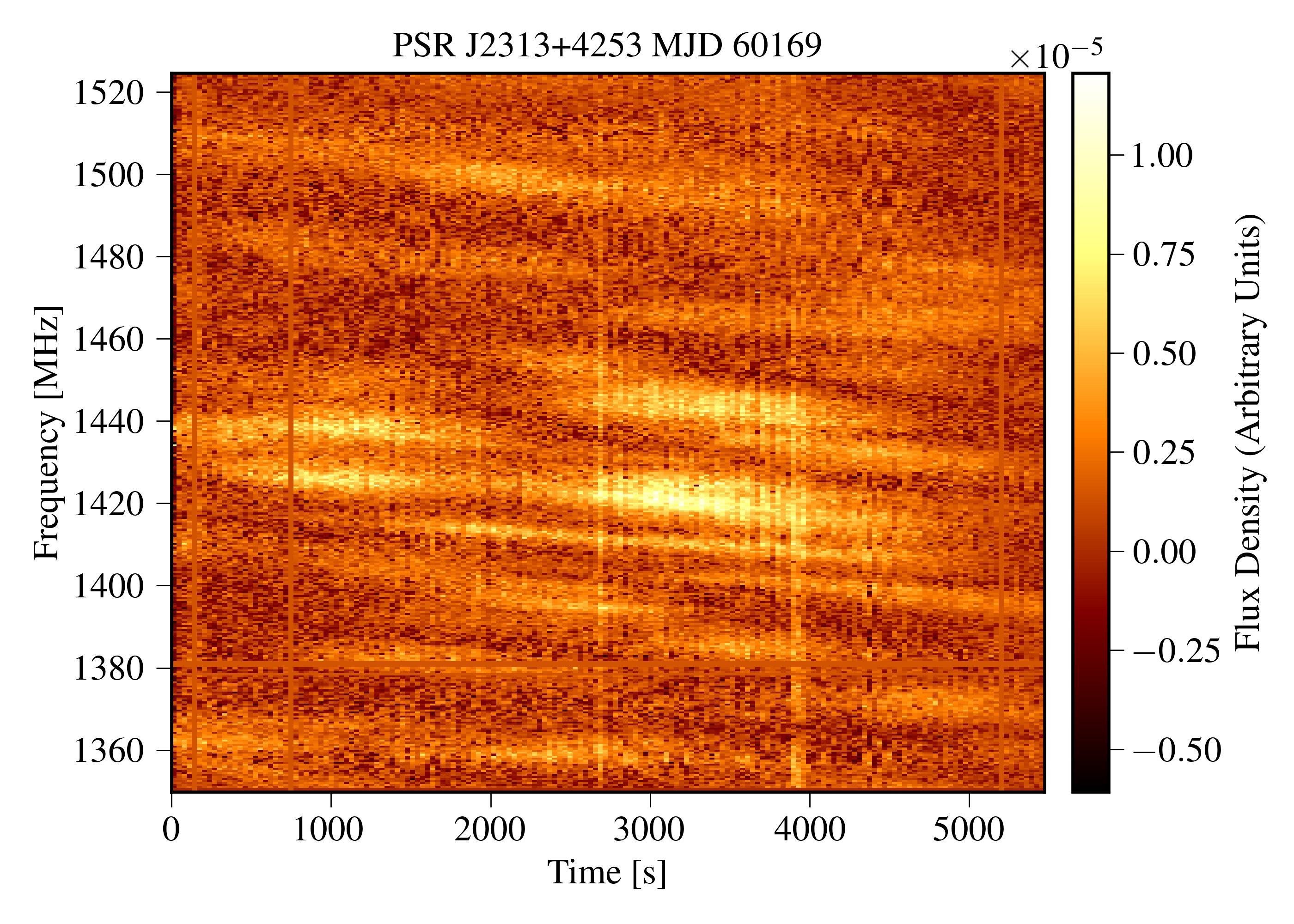}
\caption{Example dynamic spectrum for PSR J2313+4253 on MJD 60169.}
\label{2313_dyn_ex}
\end{figure}

\begin{figure}[!ht]
\centering
\hspace*{-.7cm}
\includegraphics[scale = 0.1]{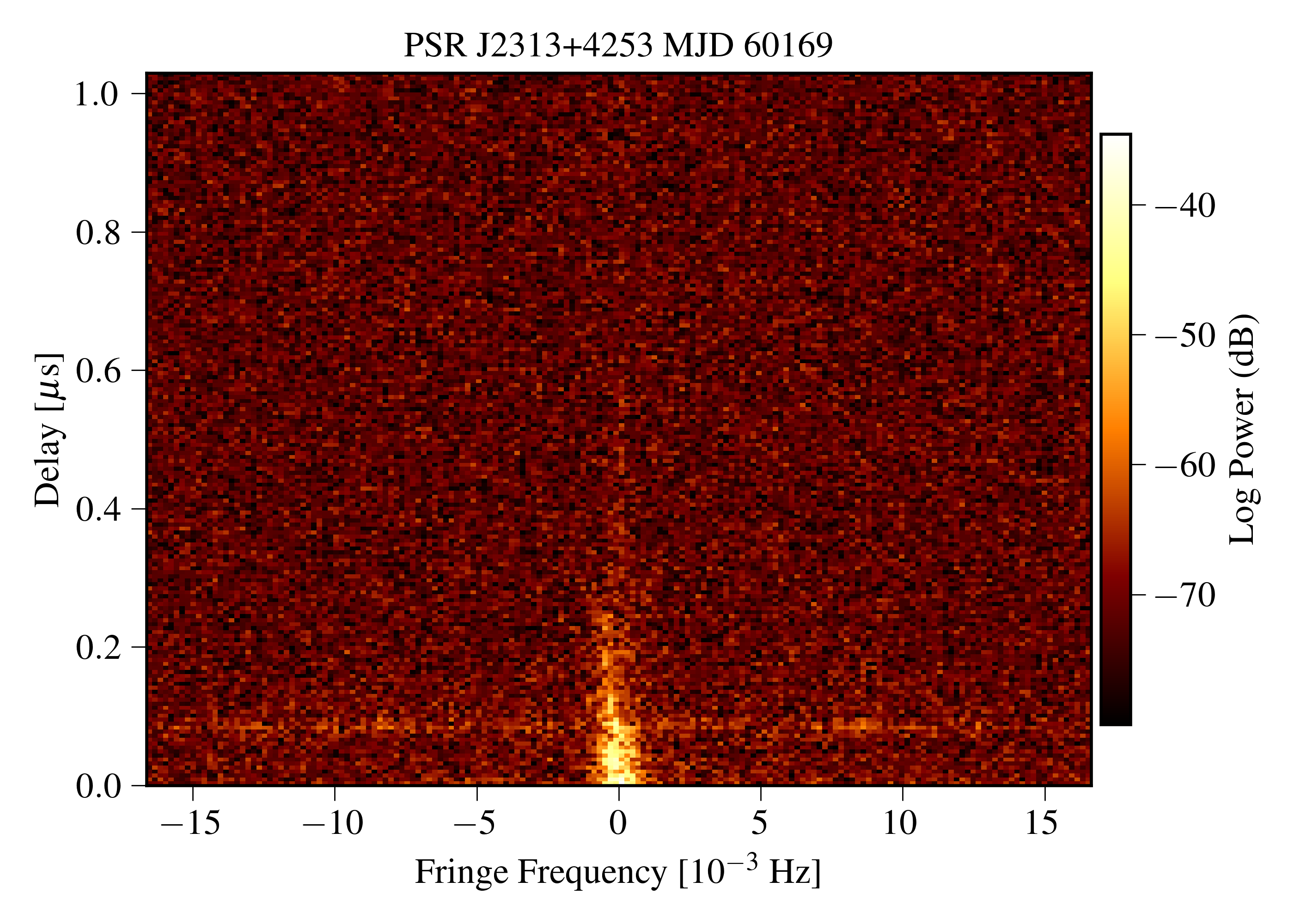}
\caption{Example secondary spectrum for PSR J2313+4253 on MJD 60169.}
\label{2313_sec_ex}
\end{figure}

\begin{figure*}[!ht]
    \subfloat[\centering Full-galaxy view showing the position of our solar system and the potential location of the scattering screen.]{\hspace*{-.45cm} {\includegraphics[width=0.55\linewidth]{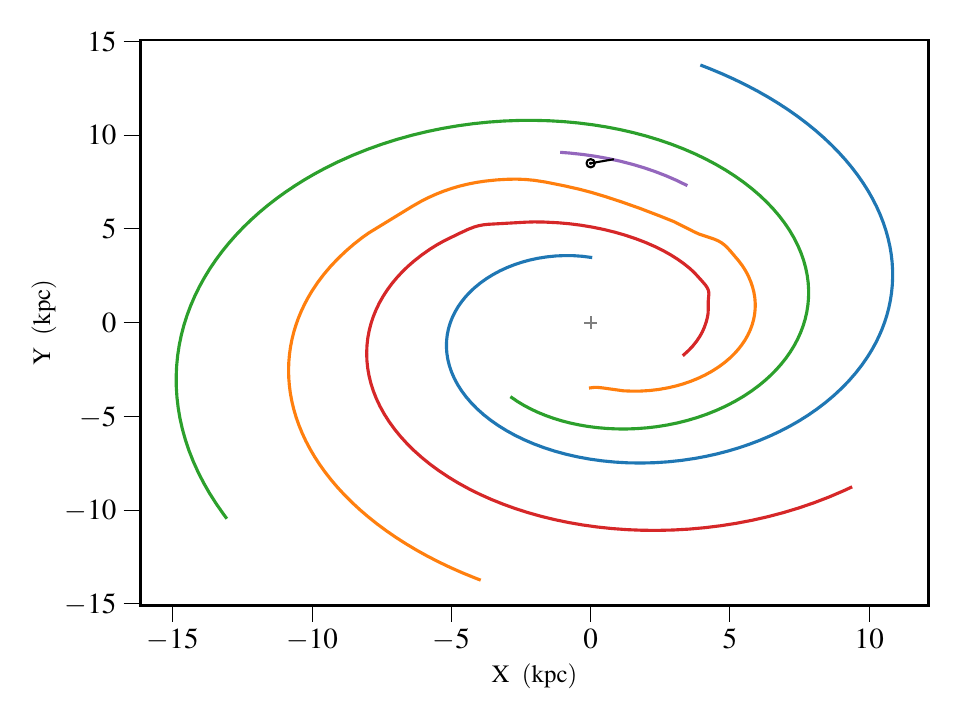} }}%
    \subfloat[A zoomed-in version of the figure on the left.]{\hspace*{-.4cm} {\includegraphics[width=.55\textwidth]{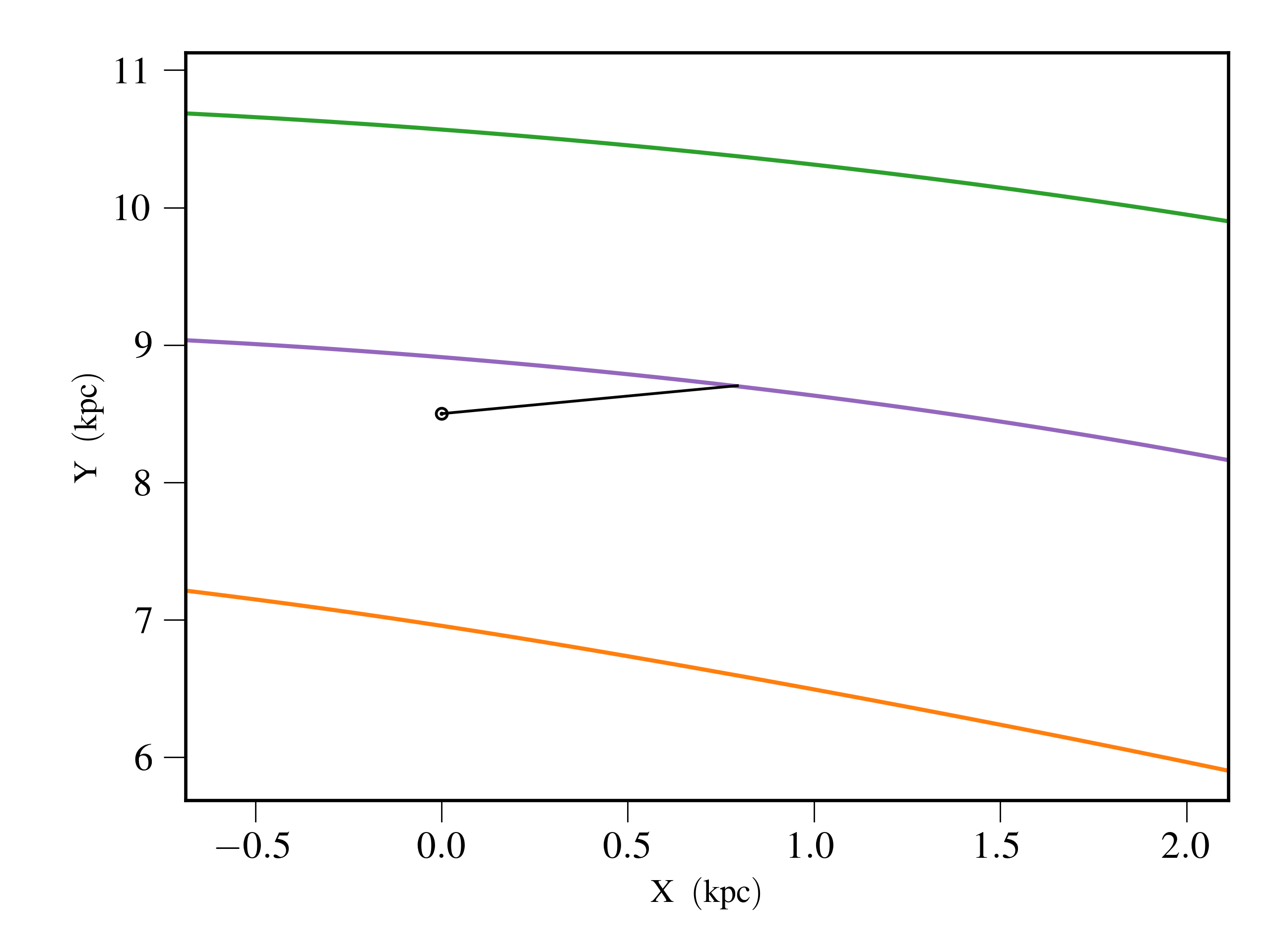} }}%
    \caption{Illustration of the arms of the Milky Way produced by the electron density model NE2001 \citep{Ocker_2024}, with the dot representing the location of our solar system and the line extending to the potential scattering screen distance of PSR J2313+4253 within the error of our estimations.}%
    \label{ne2001_galaxy}%
\end{figure*}

An example dynamic and secondary spectrum can be seen in Figures \ref{2313_dyn_ex} and \ref{2313_sec_ex}, respectively.
This pulsar is known to have at least two visible scintillation arcs \citep{stine_survey}. We resolve one arc in our own data, whose curvature matches up quite well with the arc observed at 1400 MHz by \cite{stine_survey}. Based on our transverse velocity and scintillation arc analyses, both the scintles in our dynamic spectra and scintillation arcs in our secondary spectra seem to originate from the same scattering screen, which we place around 0.8$-$1.1 kpc from Earth depending on the analysis used. Assuming our measured screen distances are not lower limits, this very likely places the screen somewhere within the Orion-Cygnus arm of our galaxy, as shown in Figure \ref{ne2001_galaxy}. This is further substantiated by the most up-to-date measurements of this pulsar's distance placing it within the Galactic plane \citep{Chatterjee_2009}. Additionally, this spiral arm is approximately 200 pc thick \citep{arm_thickness}, further expanding the range of distances that would correspond with a screen originating in this arm. Future high-precision efforts employing scintillometry will be required to confirm this screen origin, as well as whether the screen can be attributed to any discrete structure within the arm.


\section{Conclusions \& Future Work}\label{concl}
We performed high-cadence scintillation studies of eight canonical pulsars using observations taken with the Green Bank Observatory's 20m telescope. By observing each pulsar at least once a month, we obtained detailed and expansive timeseries of scintillation bandwidth and timescale and arc curvature spanning anywhere from one to three years depending on the pulsar. 
\par While the scintillation behavior of most of our pulsars appears fairly consistent across the past two decades, we find a significant change in both the measured scintillation parameters and the scintillation pattern of PSR J2022$+$5154, which we suggest may be the consequence of another scattering screen dominating the scintillation pattern.
\par We augmented results from \cite{turner_scat} showing while there exists a strong correlation between estimated scattering delays and those predicted by the NE2001 electron density model in measurements that are $7-11$ years older than our data, newer estimations of scattering delays from pulsars which were likely used in the development of this model show noticeable departure from this strong correlation. While this may partially be the result of a small sample size of pulsars, it may also provide evidence of  the ISM evolving enough on longer timescales that an update to this model may be important. We also augmented the findings of \cite{turner_scat} suggesting a positive relation between a pulsar's dispersion measure and the variability of its scattering delay over time. While, averaging over more free electrons at a higher dispersion measure might lead one to expect less scattering variability, it may be the greater number of free electrons along the line of sight may mean more opportunities for electron density fluctuations from epoch to epoch, resulting in a larger degree of variability in measured scattering delays. This finding is of particular importance to pulsar timing efforts, especially when exploring methods to correct for time-evolving scattering delays in pulsar timing.
\par Multiple pulsars exhibit evolving asymmetries within the arms of their scintillation arcs, while some arcs in PSR J1645--0317 may contain arclets. We find evidence for a previously unreported screen in one pulsar, as well as evidence that one of the screens in PSR J2313$+$4253 may lie somewhere within the Orion-Cygnus arm of the Milky Way. 
\par Future efforts will include the continued observations of the pulsars used in this study, both using the 20m telescope and the Green Bank Telescope, as well as the expansion of our campaign to include additional pulsars.

\par \textit{Acknowledgements}: We gratefully acknowledge support of this effort from the NSF Physics Frontiers Center grants 1430284 and 2020265 to NANOGrav.
\par The majority of data processing for this work took place on the NANOGrav notebook server, run by West Virginia University.

\par \textit{Author contributions}: JET led the research effort, performed data interpretation, provided guidance on the writing of Jupyter notebooks used in the analysis, and wrote the paper. JGLM wrote the main data-processing notebook and scintillation arc notebook, trained students how to use the Jupyter notebooks, and observed and analyzed the data for PSRs J2022+2854, and J2022+5154. ZZ wrote the weighted average notebook, created the pulsar screen distance visualization, and observed and analyzed the data for PSR J2313+4253. KAG observed and analyzed the data for PSR J1645$-$0317. JM assisted in analyzing the data for PSR J0332+5434. MK observed and analyzed data for PSR J0332+5434. LDCV observed and analyzed data for PSR J0922+0638. AL observed and analyzed data for PSR J2018+2839. CBF observed and analyzed data for PSRs  J0826+2637 and J2022+5154. 
\par \textit{Software}: \textsc{psrchive} \cite{2004PASA}, \textsc{pypulse} \citep{pypulse}, \textsc{scintools} \citep{Reardon_2020}, \textsc{scipy} \cite{scipy}, \textsc{numpy} \cite{numpy}, and \textsc{matplotlib} \cite{matplotlib}.
\bibliography{12_5_yr_Scattering_Paper_Draft.bib}{}
\bibliographystyle{aasjournal}
\end{document}